\documentclass[sigconf,nonacm]{acmart}

\usepackage{booktabs}
\usepackage{subcaption}
\usepackage{amsmath}
\usepackage{pifont}  

\definecolor{zoneA}{HTML}{4A65AE}
\definecolor{zoneB}{HTML}{3E8E6E}
\definecolor{zoneC}{HTML}{C46A3B}
\newcommand{\zchip}[2]{\begingroup\setlength{\fboxsep}{1.5pt}\colorbox{#1!15}{\textcolor{#1}{\textbf{\small #2}}}\endgroup}
\AtBeginDocument{%
  }

\setcopyright{none}

\begin{document}

\title[Task-State Governance: A Virtual Machine for Verifiable and Recoverable GUI Agent Execution]{Task-State Governance: A Virtual Machine for Verifiable and Recoverable GUI Agent Execution}

\author{Wenkui Yang}
\email{yangwenkui.03@gmail.com}
\affiliation{%
  \institution{SAIS, University of Chinese Academy of Sciences}
  \city{Beijing}
  \country{China}
}

\author{Yiming Hao}
\affiliation{%
  \institution{Institute of Automation, Chinese Academy of Sciences}
  \city{Beijing}
  \country{China}
}

\author{Kai Zhou}
\affiliation{%
  \institution{Meituan}
  \city{Beijing}
  \country{China}
}

\author{Chengyu Pu}
\affiliation{%
  \institution{SAIS, University of Chinese Academy of Sciences}
  \city{Beijing}
  \country{China}
}

\author{Yuzhe Zhang}
\affiliation{%
  \institution{Peking University}
  \city{Beijing}
  \country{China}
}

\author{Jie Cao}
\affiliation{%
  \institution{Institute of Automation, Chinese Academy of Sciences}
  \city{Beijing}
  \country{China}
}

\author{Jin Huang}
\affiliation{%
  \institution{Institute of Software, Chinese Academy of Sciences}
  \city{Beijing}
  \country{China}
}

\author{Shuai Ma}
\authornote{Correspondence: \texttt{mashuai@iscas.ac.cn} (Shuai Ma) and \texttt{tianfeng@iscas.ac.cn} (Feng Tian).}
\email{mashuai@iscas.ac.cn}
\affiliation{%
  \institution{Institute of Software, Chinese Academy of Sciences}
  \city{Beijing}
  \country{China}
}

\author{Feng Tian}
\authornotemark[1]
\email{tianfeng@iscas.ac.cn}
\affiliation{%
  \institution{Institute of Software, Chinese Academy of Sciences}
  \city{Beijing}
  \country{China}
}


\begin{abstract}
Graphical user interface (GUI) agents now execute workflows across real applications, but when execution drifts, users must determine what changed, what should have remained untouched, and what can still be recovered.
Plan-centric interfaces leave users supervising procedure, but users live with changes to application state.
We introduce \textbf{task-state governance}, an interaction paradigm that represents a task as an editable contract linking desired outcomes, facts that must be preserved, and observed application state.
We instantiate this paradigm in \textbf{TaskVM}, which projects live state into editable task controls, executes user edits as genuine interface actions, verifies outcomes through independent observation, and marks each action as reversible, compensatable, or irreversible.
Across an automated contract benchmark, a controlled user study, and an open live-model exploration, the paradigm strengthened contract compliance and users' recovery control and trust, while execution outcomes were held constant.
Task contracts open a design space for governable agent interfaces.
\end{abstract}



\begin{teaserfigure}
\centering
\includegraphics[width=0.84\textwidth]{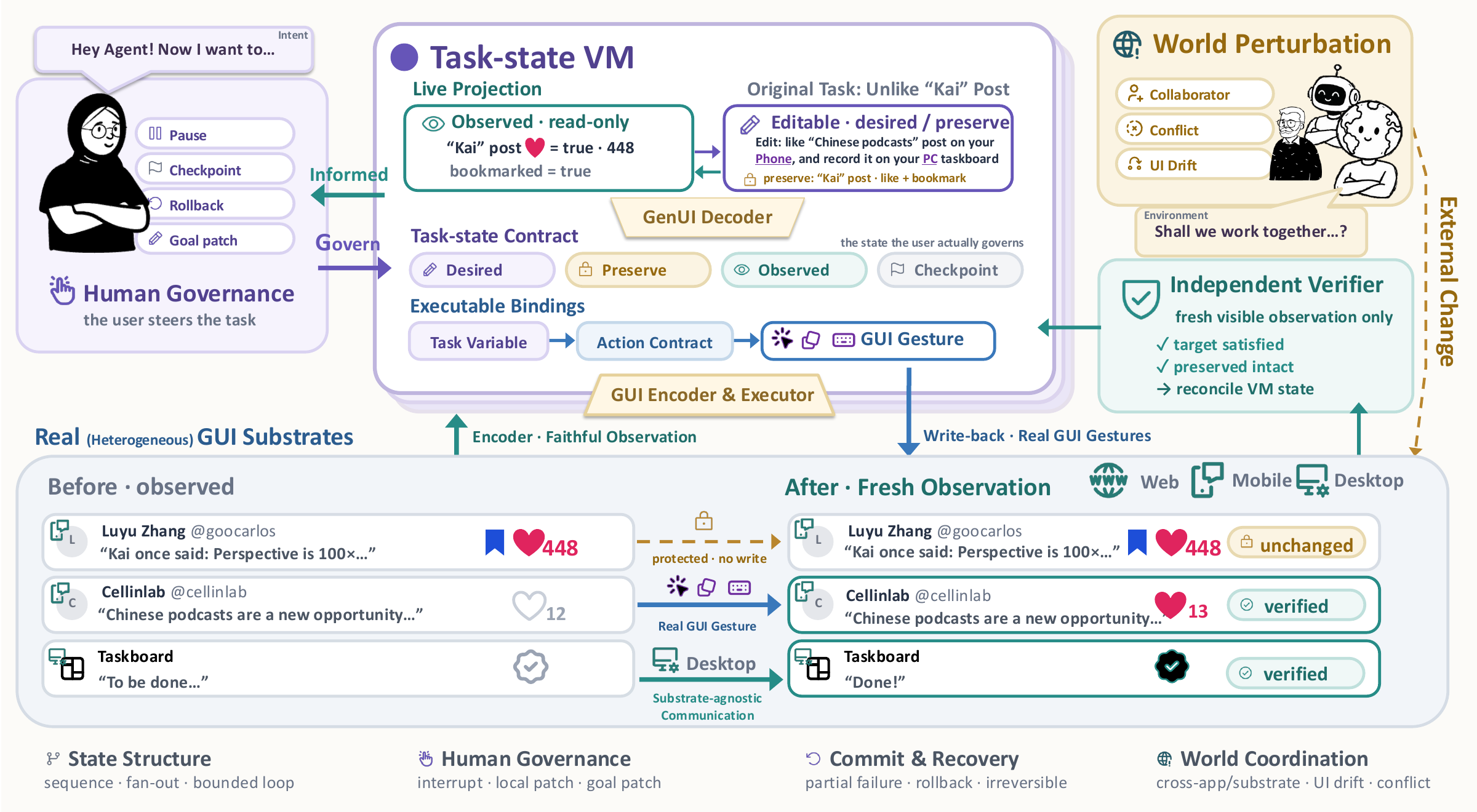}
\caption{Task-state governance on a live multi-application workflow. Rather than supervising an agent's low-level execution trajectory, the user governs a unified Task-state VM. Observed application state (bottom left) is projected into interactive task variables (desired, preserve, observed); edits are translated into real GUI gestures by the GUI Encoder and Executor, while an Independent Verifier checks fresh visible observations against non-interference constraints. Image is a semantically translated replica of the production interface.}
\label{fig:teaser}
\end{teaserfigure}

\maketitle


\section{Introduction}

Consider postponing a project briefing by one day.
The calendar entry and the team announcement must change, while a settled flight booking must remain untouched, and a collaborator may request a different date midway through execution.
A revised plan does not, by itself, say which updates have already taken effect, which records must stay protected, and which effects can no longer be undone.
As Figure~\ref{fig:teaser} illustrates, users ultimately live with changes to the world, not with changes to an agent's plan.

Computer-use agents (CUAs) have made this delegation technically plausible at scale.
Modern agents parse complex layouts, synthesize input gestures, and execute extended operational sequences across unfamiliar interfaces \citep{openai2025cua, cheng2024seeclick, lin2025showui, zheng2024seeact, deng2023mind2web, he2024webvoyager, rawles2025androidworld}.
Interaction research has responded by opening execution up for oversight: co-planning surfaces let users revise procedural steps, ambient displays communicate agent status, and malleable or generative interfaces reorganize the workspace around the task \citep{feng2026cocoa, huq2025cowpilot, venkatesan2026plover, chang2026sidekick, kim2026agentlens, jiang2026orca, xu2026duetui, cao2025jelly}.
These designs make the agent's procedure the object of interaction: what the user inspects and edits is what the agent will attempt next.

Two gaps persist in this procedure-centric stance.
First, editing a step changes what the agent will attempt, but it cannot express what the world must keep: boundaries on side effects, records that must remain untouched, and facts that should outlive the current plan have no representation in a list of actions.
Second, completion and recovery rest on the agent's own account.
When the executor's report is the only evidence, users cannot verify what actually changed, and undo is offered without a model of which effects can be restored, inviting the over-reliance that supervisory control research has long documented \citep{bainbridge1983ironies, parasuraman1997misuse, parasuraman2010bias, lee2004trust}.
Both gaps follow from one representational choice: the task is modeled as a sequence of actions, so the obligations attached to its outcome have nowhere to live.

Our stance extends a lineage of work that treats the object, not the procedure, as the unit of interaction.
Direct manipulation established that users reason about visible objects of interest \citep{shneiderman1983direct, hutchins1985direct}; undo research made recovery a first-class design concern \citep{abowd1992undo, myers1996commands}; and accountability research argued that systems should expose what they did so that people can judge it \citep{bellotti2001intelligibility, suchman1995visible}.
We extend this lineage from artifacts the user operates to agents that operate on the user's behalf.
We call the resulting stance task-state governance: a task is represented as an editable contract among three state components, not as a procedure alone.
The \textbf{desired state} holds the target values the user authorizes the agent to establish.
The \textbf{preservation scope} pins application facts that must remain undisturbed.
The \textbf{observed state} captures what running applications currently display.
When the three are represented together, an edit becomes a checkable commitment: the system knows what it must establish, what it must not disturb, and what evidence will count as done.
The plan remains useful, but as one revisable realization of the contract rather than as the task itself.

We call the runtime that maintains this contract a \textbf{Task Virtual Machine}, and the name is deliberate about scope.
It is not an operating-system virtual machine: it owns neither the full state of third-party applications nor the power to roll them back atomically.
What it owns is the task's obligation state and an honest account of evidence.
Four commitments make the abstraction tangible.
Editing a desired value does not assert that the world changed; desired and observed values sit side by side until fresh observation closes the gap.
Marking a record as preserved lifts that constraint out of the agent's plan and into independent verification.
An executor's completion report does not close a task; a separate verifier checks fresh screen evidence first.
And a checkpoint is not time travel; recovery is expressed as the operations the environment actually supports.

We realize this stance in \textbf{TaskVM}, a prototype that places the observed state of live applications beside editable task controls.
The runtime compiles user edits into genuine interface gestures, so authority flows through the same rendered interface a human would use.
An \textbf{independent visual verifier}, architecturally separated from the executor, checks fresh observations against the contract, and the interface distinguishes reversible, compensatable, and irreversible actions so that recovery is described as supported operations rather than a blanket promise.
As the Java Virtual Machine (JVM) exposes one bytecode interface across operating systems, the task virtual machine exposes one obligation model across software stacks; substrate independence is the extent of the analogy we intend.

We evaluate the paradigm through three complementary lenses, asking three research questions (RQ):
\begin{itemize}
\item \textbf{RQ1:} Does representing tasks as state contracts improve machine-level satisfaction of governance obligations compared with procedural agent baselines, and at what execution cost?
\item \textbf{RQ2:} When execution outcomes are held constant, does governing state change how well users understand, control, and trust the agent?
\item \textbf{RQ3:} Does the paradigm remain usable and welcome when a live model executes tasks that participants author themselves?
\end{itemize}

An automated benchmark over 36 task instances drawn from a suite of productivity, social, and content applications measured contract satisfaction and its costs.
A controlled within-subject study with 37 participants compared TaskVM against a conversational agent interface under replayed execution, isolating perception from execution variance.
An exploratory follow-up with 14 returning participants observed live-model use on self-authored tasks.
Across these lenses, TaskVM improved contract compliance at higher machine cost, significantly increased perceived recovery control and trust while execution outcomes were held constant, and remained usable in open-ended use, where participants exercised recovery operations and asked for more.

In summary, this paper contributes:
\begin{itemize}
\item \textbf{Task-state governance}, an interaction model that represents a task as an editable contract over desired outcomes, preservation constraints, and observed state, together with an explicit taxonomy of recovery.
\item \textbf{TaskVM}, a prototype that instantiates the model through live state projection, GUI-grounded bidirectional execution, independent verification, and honestly bounded recovery, running across web, mobile, and desktop substrates.
\item \textbf{A multi-tier evaluation} combining an automated contract benchmark, a controlled user study, and an open live-model exploration, together with design implications for agent interfaces that users can govern rather than merely watch.
\end{itemize}


\section{Related Work}

Our research intersects four traditions: autonomous graphical user interface agents, interfaces that project agent execution, bidirectional and malleable data abstractions, and human-agent governance and awareness.
Appendix Table~\ref{tab:related-matrix} maps this landscape along the object each interface asks the user to govern, tracing what an edit commits each system to, what counts as completion, and what recovery can mean.

\subsection{GUI Agents: From Autonomous Execution to Interaction}

Autonomous graphical user interface agents have transitioned from scripted web automation to multimodal models operating across web \citep{zhou2024webarena, koh2024visualwebarena, deng2023mind2web}, desktop \citep{xie2024osworld, hu2025osagents}, and mobile operating systems \citep{rawles2025androidworld, wu2026mobilegym, zhang2024mobileenv}.
Modern systems specialize in coordinate-level display grounding \citep{cheng2024seeclick, gou2025uground, xu2025aguvis, lin2025showui, qin2025uitars}, supported by structured reasoning loops \citep{yao2023react, zheng2024seeact, he2024webvoyager}.
However, benchmark success on autonomous execution does not resolve the human interaction problem.
Autonomous traces remain opaque, self-reported completion often diverges from application reality, and single-pass metrics ignore the interactive cost of correcting intermediate errors.
Rather than proposing another grounding model, we investigate how agents operate within an architecture that enforces state transparency, non-interference, and verifiable outcomes.

\subsection{Projecting Agent Execution: Plans, Interfaces, and Malleable Views}

To support human oversight, prior research has explored diverse representations of agent operations.
Interactive scaffolding demonstrated the value of structured prompt chains for inspectable sub-steps \citep{wu2022chains}.
Plan-centric systems externalize reasoning into sequential steps, decision forks, and progress indicators \citep{feng2026cocoa, huq2025cowpilot, mozannar2025magentic}.
Plover represents the closest point of comparison: it surfaces structured plans that users can locally revise, keeps executed plan steps as immutable history, attaches derived constraints and visual checks to actions, and runs a cross-platform executor across desktop environments \citep{venkatesan2026plover}.
Beyond a six-person formative study, its evaluation includes an expert benchmark-repair analysis in which the first author intervened on 26 autonomous failures and improved 23, presented by the authors as an upper bound on recoverability rather than typical user performance.

Our comparison concerns the object of interaction and the responsibilities attached to an edit.
In plan-centric designs, the protected object is the plan's executed history, whereas the obligations we study protect facts inside external applications.
The evidence of completion also differs in scope: plan-centric systems inspect plan state and screens for execution continuity, whereas our verifier certifies explicit contract obligations (target satisfaction and non-interference over protected facts) from fresh, independently obtained observations of application state.
And the evaluative claims attach to different populations: expert-assisted recoverability in a formative setting, versus unaided use by ordinary participants under controlled replay.
We read these framings as protecting different assets, and the two objects can compose within one system.

Complementary work explores malleable views and generative interfaces.
Morae pauses browser workflows to preserve critical user choices \citep{peng2025morae}.
Sidekick visualizes multi-task status \citep{chang2026sidekick}, while AgentLens adaptively renders mobile interfaces \citep{kim2026agentlens}.
In web mashups, Orca extracts components with bidirectional DOM synchronization \citep{jiang2026orca}, and A-mash combines mobile features into task mashups \citep{lee2022amash}.
Generative UI frameworks synthesize on-demand task cards \citep{xu2026duetui, cao2025jelly, xie2026sac, kong2026macaron, vaithilingam2024dynavis}.
Systems such as DynaVis, which composes widgets on demand \citep{vaithilingam2024dynavis}, and Orca, which organizes web components across pages \citep{jiang2026orca}, anchor their evidence loops inside the browser they render; cross-application compensation and fresh-observation checks of protected state lie outside the problems these systems set out to solve.

\subsection{Editable and Bidirectional Interfaces: Lenses, Sagas, and Views}

Decoupling physical storage from logical views originates in relational databases \citep{codd1970relational} and declarative querying \citep{chamberlin1974sequel}.
Translating edits on projected views back into source structures is formalized as the view-update problem, addressed by Foster et al. through bidirectional lenses with round-trip fidelity \citep{foster2007lenses}.
In distributed computing, managing multi-step mutations led Garcia-Molina and Salem to formulate Saga transactions with compensating steps \citep{garciamolina1987sagas, gray1981transaction, helland2007life}.

In HCI, declarative generation decouples task specifications from widget layouts, as demonstrated by SUPPLE \citep{gajos2004supple}.
Recent declarative systems like DMI \citep{wang2026dmi} and VeriSafe \citep{lee2025verisafe} compile state targets and safety constraints into accessibility gestures.
Generative UI protocols such as A2UI \citep{google2026a2ui} and GenUI \citep{leviathan2026genui} define standards for streaming components, while UICoder \citep{wu2024uicoder} and Cells \citep{kim2023cells} present methods for interactive UI code generation and structured interfaces.
TaskVM adapts these principles to computer-use agents, treating heterogeneous interfaces as a distributed database and compiling user edits into verified, compensatable actions across applications.

\subsection{Governance, Trust, and Awareness in Human-Agent Collaboration}

Designing effective oversight requires mitigating automation complacency.
Human supervisory performance degrades during passive monitoring \citep{bainbridge1983ironies}, fostering automation bias when agents exhibit high but imperfect competence \citep{parasuraman1997misuse, parasuraman2010bias}.
Mixed-initiative frameworks advocate dynamic responsibility allocation \citep{horvitz1999mixed, maes1994agents}.
Human-AI guidelines emphasize communicating system actions, supporting efficient error correction, and conveying operational boundaries \citep{amershi2019guidelines}.
Furthermore, mitigating over-reliance requires cognitive forcing functions that prompt deliberate verification \citep{lee2004trust, hoff2015trust, bucinca2021trust}.

Human agency is further supported by malleable, reversible interaction.
HCI inquiries show that users continuously appropriate software to fit evolving practices \citep{dourish2003appropriation, dix2007appropriation, tchounikine2017appropriation}.
Reversibility underpins direct manipulation through rapid, reversible operations with visible impact \citep{shneiderman1983direct, hutchins1985direct}, formalized in command trees, selective undo, and side views \citep{abowd1992undo, myers1996commands, terry2002sideviews}.
Across multi-device and task-centric computing, systems maintain continuity across fragmented tools \citep{rekimoto1997pickdrop, rekimoto1999surfaces, brudy2019crossdevice, henderson1986rooms, kaptelinin2003umea, bardram2005activity, lachand2019toccata}.
Finally, collaborative awareness requires visible artifacts that expose system capabilities and boundaries \citep{dourish1992awareness, heath1992control, suchman1995visible, bellotti2001intelligibility, kulesza2015debugging, sengers2005reflective}.
TaskVM synthesizes these insights by replacing passive trajectory tracking with an editable, live projection of verified application state.


\begin{figure*}[t]
\centering
\includegraphics[width=\textwidth]{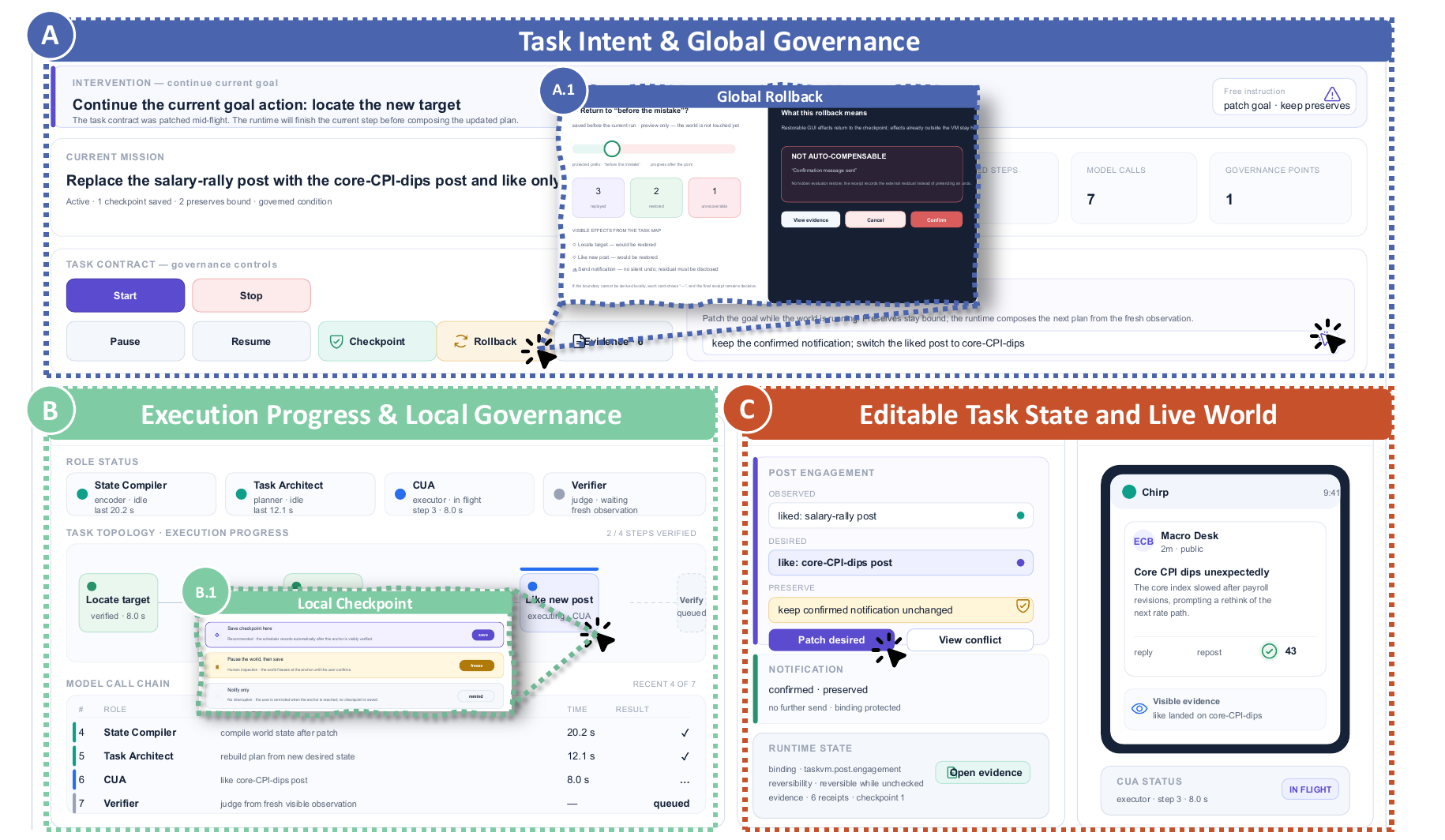}
\caption{The TaskVM governance interface (a faithful English rendering of the production UI, redrawn for language accessibility). \zchip{zoneA}{A} Task Intent \& Global Governance: the current mission and task contract, lifecycle controls (start, pause, resume, checkpoint, rollback), and a global rollback dialog that previews which committed steps would be replayed or restored and flags effects that are not auto-compensable. \zchip{zoneB}{B} Execution Progress \& Local Governance: live status of the four model-invoking roles (state compiler, task architect, CUA executor, verifier), task-topology progress, local checkpoint options, and the model call chain. \zchip{zoneC}{C} Editable Task State \& Live World: the desired--preserve--observed triad rendered beside the live GUI, with patch and conflict controls and a runtime-state card linking every variable to visible evidence.}
\label{fig:interface}
\end{figure*}

\section{A Task-State Governance Interaction Model}

To bridge human intent and autonomous software execution, we formulate a task-state governance model elevating application state into the primary interactive medium: humans govern tasks, while agents operate applications.

\subsection{State Intent versus Plan Intent}

Current interactive agents structure human intervention around plan intent, presenting upcoming procedural steps, tool invocations, or sub-goal trees \citep{feng2026cocoa, venkatesan2026plover}.
When a user intervenes, the operation modifies an upcoming action, such as replacing a query or deleting a click.
While offering procedural steerability, this conflates what the agent attempts with what the user requires the world to become: a step can execute without error yet fail to achieve the intended outcome, or inadvertently corrupt adjacent records \citep{bainbridge1983ironies, amershi2019guidelines}.

In contrast, our interaction model formalizes user intent as state intent, specifying invariant properties that must hold true once execution concludes.
Instead of dictating navigation paths, the user specifies what application facts must change and what historical context must remain undisturbed.
As defined in our domain schema, an \texttt{ActionContract} encapsulates these state obligations independently of platform execution paths.
This distinction decouples human governance from low-level software idiosyncrasies: just as programming runtimes allow developers to declare invariants while delegating memory management, our model enables users to express task invariants while delegating mouse coordinates and Document Object Model (DOM) traversals to autonomous executors.
As mapped across the top band of Figure~\ref{fig:arch}, a governed round trip traverses eight semantic stages from Observe and Abstract through Reconcile to close a single episode.

\subsection{The Executable View: Bottom-Up Live Projection}

To make state intent manipulable without exposing raw complexity, our model compiles live software state into an executable view, compressing distributed fields into a unified, low-cognitive-load surface.
Figure~\ref{fig:interface} shows the governance interface that realizes this view, organized into three zones that mirror the state representations of our model.

Zone~\zchip{zoneC}{C} (Editable Task State and Live World) presents the desired, preserve, and observed fields of each task variable side by side with the live application window.
The observed column is a read-only faithful projection of current software reality, grounded in visual evidence parsed directly from rendered screens \citep{suchman1995visible}; the desired column is editable, and the preserve column pins facts that must remain undisturbed.
Zone~\zchip{zoneA}{A} (Task Intent and Global Governance) hosts the mission contract and lifecycle controls, and Zone~\zchip{zoneB}{B} (Execution Progress and Local Governance) exposes execution progress, the live status of each model-invoking role, and the model call chain.

Maintaining fidelity between the projection and external software requires continuous bidirectional synchronization.
The projection streams dynamic live deltas through server-sent events, updating the interface as the external world changes.
When concurrent external modifications occur, such as a colleague editing a shared document, the system triggers reconciliation mechanisms.
Rather than silently overwriting external modifications or crashing the pipeline, TaskVM surfaces external changes as explicit conflict markers in Zone~\zchip{zoneC}{C}, inviting the user to reconcile divergent values through deliberate interaction \citep{dourish1992awareness, bucinca2021trust}.

\subsection{The Desired-Preservation-Observed Triad}

We formalize task-state governance through three interconnected state representations:
\begin{equation}
\mathcal{T} = \langle S_{\text{desired}}, S_{\text{preserve}}, S_{\text{observed}} \rangle
\end{equation}
The desired state $S_{\text{desired}}$ consists of target variables $\{v_1 \mapsto c_1, \dots, v_n \mapsto c_n\}$ that the user authorizes the agent to establish.
The preservation scope $S_{\text{preserve}}$ defines invariant constraints over protected application entities.
In real-world workflows, accomplishing a goal often threatens adjacent data; for example, updating a meeting time must not alter room reservations or attendee notes.
By formalizing $S_{\text{preserve}}$ as an explicit mathematical construct, our model establishes verifiable non-interference boundaries: any execution establishing $S_{\text{desired}}$ while violating an invariant in $S_{\text{preserve}}$ is classified as an operational failure.
The observed state $S_{\text{observed}}$ captures visual facts parsed from rendered application screens at time $t$.
A task transition is formally valid if and only if fresh visual observation confirms that external software satisfies the desired state while leaving the preservation scope inviolate:
\begin{equation}
\text{Valid}(\mathcal{T}) \iff (S_{\text{observed}} \models S_{\text{desired}}) \land (S_{\text{observed}} \models S_{\text{preserve}})
\end{equation}
The user governs the relationship between $S_{\text{desired}}$ and $S_{\text{preserve}}$ in the governance console (Zone~\zchip{zoneC}{C} of Figure~\ref{fig:interface}), while the runtime engine continuously reconciles $S_{\text{observed}}$ against external software.

\subsection{Verification Semantics: Independent Witnessing and Non-Interference}

A core tenet of our governance model is that execution completion must never be self-certified by the executing agent.
Autonomous agents frequently report success prematurely, mistaking the emission of a completion token for actual state alteration \citep{zheng2024seeact, he2024webvoyager}.
Our model delegates verification to an independent visual verifier evaluating rendered application screens.

The verification engine enforces two complementary properties: target satisfaction and non-interference.
To verify target satisfaction, the verifier requires visible witness predicates, demanding explicit evidence that an intended modification was actively enacted to eliminate the no-op loophole where an agent claims success on a pre-existing state.
To enforce non-interference, the verifier evaluates negative predicates over specified, observable entities in $S_{\text{preserve}}$, confirming that protected fields display the same values in fresh screenshots.
Crucially, the independent verifier operates with strict contextual and architectural isolation from the executing agent, sharing no internal reasoning buffers or prompt scaffolding; in the interface it appears as a first-class role awaiting fresh observation (Zone~\zchip{zoneB}{B} of Figure~\ref{fig:interface}).
Evaluating fresh visual observations against formal predicates provides an objective, falsifiable gate before any task is marked complete in the user interface \citep{bellotti2001intelligibility}.

\subsection{Recovery Semantics: Reversible, Compensatable, and Irreversible Actions}

Real-world software interactions require robust mechanisms for error correction, interruption, and recovery.
In traditional direct manipulation interfaces, undo operations revert local memory structures \citep{shneiderman1983direct, abowd1992undo}.
However, in multi-application environments, software changes induce persistent external side effects that cannot be undone by simple state restoration \citep{gray1981transaction, garciamolina1987sagas}.

Our model establishes a taxonomy of operational reversibility, codified in our implementation as \texttt{Reversibility}:
\begin{itemize}
\item \textbf{Reversible Actions:} Operations whose effects can be completely and symmetrically undone through standard graphical gestures without residual impact, such as toggling an unsubmitted checkbox or clearing a draft text field.
\item \textbf{Compensatable Actions:} Operations that induce persistent external effects but can be semantically counterbalanced through secondary compensation actions, such as sending a cancellation message to follow an erroneous notification \citep{garciamolina1987sagas, helland2007life}.
\item \textbf{Irreversible Actions:} Operations that permanently modify external systems or invoke physical effects that cannot be restored, such as deleting account archives or completing financial settlements.
\end{itemize}
When a user initiates a rollback or edits a historical checkpoint, TaskVM derives a compensation plan exclusively from its committed action history (CompensationPatch).
Reversible and compensatable actions execute compensating GUI gestures across affected applications.
Conversely, when an action crosses an irreversible boundary, the interaction model prohibits deceptive representations of undo: the global rollback dialog in Zone~\zchip{zoneA}{A} (Figure~\ref{fig:interface}) previews which committed steps would be replayed or restored, explicitly marks effects that are not auto-compensable, and locks irreversible entities with transparent status badges, preventing the dangerous illusion of complete reversibility and supporting appropriate reliance \citep{lee2004trust, bucinca2021trust}.


\begin{figure*}[!t]
\centering
\includegraphics[width=\textwidth]{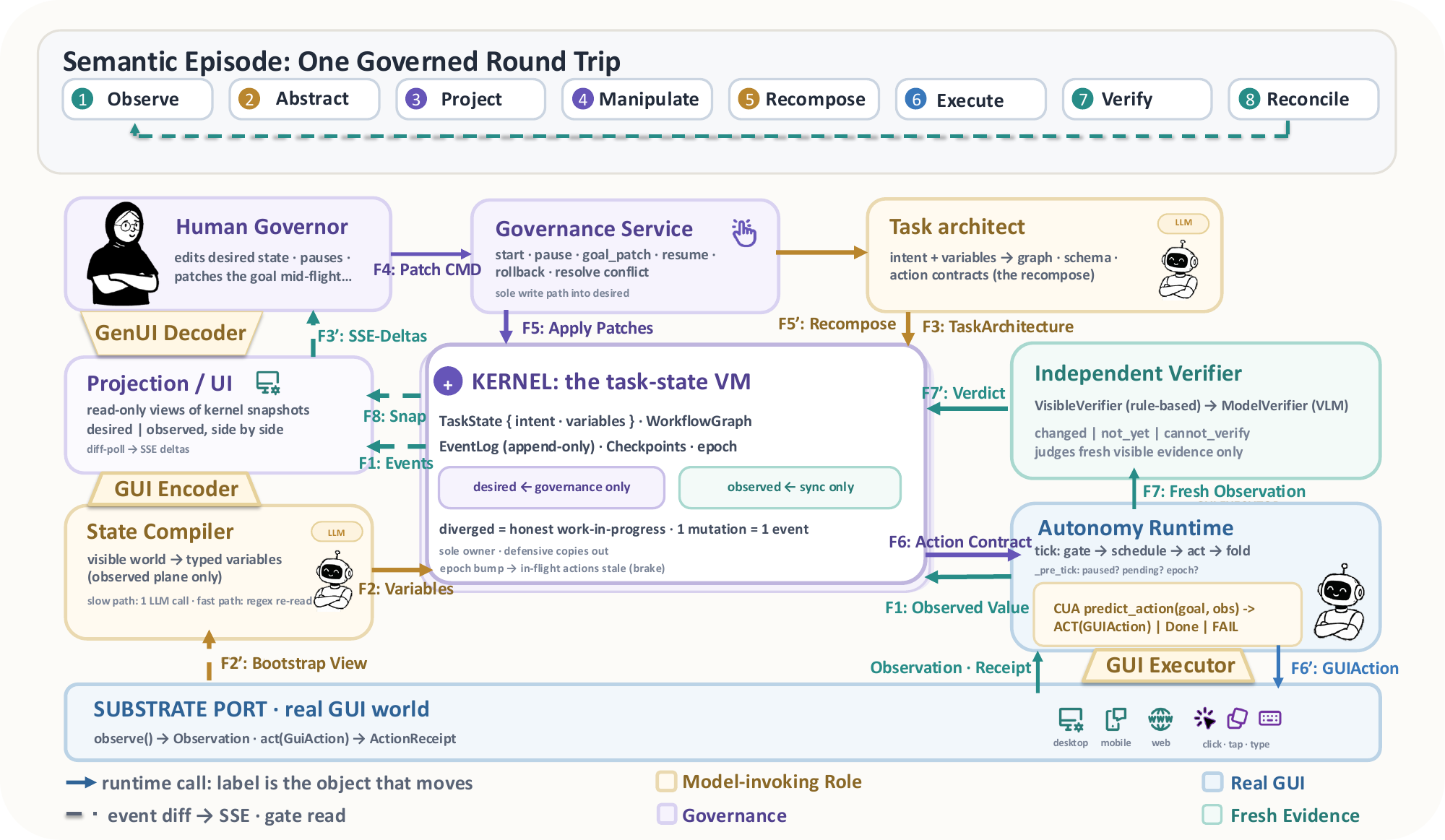}
\caption{TaskVM data flow as a hub-and-spoke architecture. The kernel is the sole owner of task state; six spokes exchange typed objects with it, and the substrate port is the only door to the real GUI world (\emph{observe() } in, \emph{act(GuiAction)} out; real gestures only, with no backdoor writes). Every arrow is labeled with the actual data object that moves. The label prefix \textbf{F1 to F8} indexes the eight semantic stages of one governed round trip (top band), so a number denotes \emph{stage membership}, not size or importance; a prime ($'$) marks the \emph{paired/return process of the same stage} rather than a different stage (for example, F7 fresh observation and F7$'$ verdict are the question and the answer of \emph{Verify}). Amber cards host model-invoking roles (architect, state compiler, CUA model); violet equals governance, teal equals fresh evidence, blue equals real GUI.}
\label{fig:arch}
\vspace*{-14pt}
\end{figure*}

\section{The TaskVM Prototype}

We implement the task-state governance model in TaskVM, an interactive system governing multi-application workflows across web, mobile, and desktop software.
This section details the data flow, privacy boundaries, bidirectional bindings, governance primitives, and substrate abstractions that realize our conceptual framework.

\subsection{System Architecture and Separation of Concerns}

As mapped in Figure~\ref{fig:arch}, TaskVM is organized around a hub-and-spoke data flow centered on the Task-state VM (Kernel).
The Kernel serves as the sole owner of task state, maintaining TaskState (intent, variables), workflow graphs, append-only event logs, and epoch checkpoints.
By enforcing a defensive read path, the system prevents low-level drivers or downstream consumers from polluting central task semantics.

Rather than relying on a single vision-language model, TaskVM distributes responsibilities across specialized roles that exchange typed objects with the Kernel.
The GUI Agent functions as an Encoder, observing live software displays via dedicated observation structures and channeling them into the State Compiler (F2).
The Compiler converts visual candidates into standardized task variables (F2, F2') and forwards them to the Task Architect.
A separate generative interface model functions as a Decoder (F3), translating intent into interactive A2UI components projected to the user interface (F3').
These roles execute through independent model invocations marked by amber indicators, preventing their conversational histories from being shared directly.
All experiments reported in this paper instantiate every model-invoking role with the same production model, OpenAI's GPT-5.6-sol (per-condition configuration in Section~\ref{sec:stage1} and Appendix~\ref{app:evalconfig}).
While this separation limits direct context leakage, it does not eliminate extraction errors or correlated failures, so the generated representation is not treated as evidence of its own correctness.

\subsection{GUI-Only Purity and the No-Internal-ID Rule}

A pervasive vulnerability in automated GUI systems is reliance on internal implementation identifiers, such as private DOM IDs, database keys, or synthetic accessibility tags.
While internal identifiers facilitate automated testing, they break down in open-world environments where dynamic layouts alter internal structures without changing visual appearances.
Furthermore, relying on internal identifiers creates backdoors that bypass rendered interfaces, giving automated agents privileges unavailable to human operators.

To maintain operational integrity, TaskVM enforces the GUI-only purity standard.
Every piece of data entering the model pipeline must satisfy the one-question test: could a real human user see this exact string on the rendered screen?
If an attribute cannot be visually perceived by a human looking at the monitor, it is strictly prohibited from entering any model prompt.
The compilation pipeline incorporates an explicit verification module that intercepts candidate records and strips internal database keys, accessibility labels, operator jargon, and deep-link URLs via dedicated hidden-ID scrubbers and architectural no-leak filters.
Every task variable in the governance console must link directly to verbatim visual evidence parsed from the display.
Moreover, all write operations must traverse authentic graphical gestures (GuiAction), including coordinate-targeted mouse clicks, keystrokes, and scroll events.
Any state alteration that does not require a rendered display and simulated user input is classified as an unauthorized backdoor and rejected by the runtime kernel.

\subsection{Bidirectional Executable Binding and Multi-App Action Contracts}

TaskVM achieves bidirectional interaction through executable semantic bindings that connect high-level virtual machine variables to low-level application gestures.
We conceptualize this mechanism as an analog to the Java Virtual Machine.
Just as a Java programmer modifies an abstract object variable and relies on the virtual machine to translate the assignment into platform-specific memory offsets, a TaskVM user edits an abstract task variable in the governance console and relies on TaskVM to synthesize coordinated writes across heterogeneous software.

When a user alters a variable in the governance console, the system instantiates a semantic \texttt{ActionContract} in our domain schema.
The contract specifies what condition must become true in the target application, rather than hardcoding how the application must be navigated.
Target entities are identified through robust visual anchors, such as surrounding label text, relative geometric offsets, and icon visual embeddings, rather than brittle pixel coordinates.
When a single conceptual edit requires modifications across multiple applications, such as synchronizing a deadline across a project management board and a team chat client, the runtime orchestrates a fan-out execution plan.
The execution engine sequences operations across application windows, monitors UI transitions, and commits state updates only after verifying structured action outcomes (ActionReceipt).

\subsection{Hot Governance: Pausing, Patches, Epochs, and Recovery}

Real-world human governance requires the ability to intervene while execution is actively underway.
TaskVM provides four foundational hot governance primitives: pause, local patch, goal patch, and rollback.
At any point during autonomous execution, the user can trigger a pause control in the governance console.
The Governance Service immediately issues a halt signal to the active runtime, freezing the execution harness at the current step boundary.

To manage race conditions during mid-execution interventions, TaskVM introduces epoch-based governance gating in the kernel.
Every task execution belongs to a monotonically increasing integer epoch.
When a user pauses the system and applies a local patch or revises the overarching goal, the governance engine increments the epoch counter.
If the autonomous agent was in the process of generating an action from the preceding epoch, the kernel gate intercepts the returning proposal and discards it before dispatch.
This prevents stale proposals from being newly dispatched after a revision, without claiming to reverse effects of actions already dispatched.

When a user modifies a goal, the system retains valid, committed action history and recomposes only the affected future trajectory, avoiding redundant re-execution of unaffected milestones.
When a rollback is requested, the compensation engine analyzes the committed action registry.
For reversible operations, the runtime synthesizes inverse graphical gestures, such as untoggling checkboxes or clearing input forms.
For compensatable operations, the runtime executes compensating GUI actions, such as dispatching cancellation notes; these counteract the intended effect but do not assure atomic rollback \citep{garciamolina1987sagas}.
Finally, when an operation crosses an irreversible threshold, such as a finalized monetary payment, the system prevents misleading undo representations by raising an explicit irreversible signal.
Irreversible entities are locked in the interface with transparent explanatory badges, clarifying the limits of automated recovery.

\subsection{Substrate Independence and Session Isolation}

A central objective of TaskVM is substrate independence: ensuring that the same high-level task semantics can be governed identically regardless of the underlying operating system or application technology stack.
TaskVM abstracts external software through a clean SubstratePort interface, implemented across modular sessions.
The SubstratePort defines primitive operations for capturing visual screenshots, reading device metadata, and dispatching standardized mouse and keyboard input events.
Our prototype provides implementations for three distinct platforms: modern web applications, the OSWorld Linux desktop benchmark environment \citep{xie2024osworld}, and the browser-hosted mobile simulation environment MobileGym \citep{wu2026mobilegym}.
Independent visual verification takes place over extracted visual artifacts.

To isolate the agent from hidden laboratory shortcuts, TaskVM enforces strict privilege separation between the runtime session and the evaluation harness.
The user-facing runtime interacts exclusively with a session object possessing only standard human capabilities: capturing visual display pixels and issuing realistic graphical gestures.
Privileged capabilities required by benchmarking, such as resetting databases, reading internal state trees, or injecting deterministic world events, are encapsulated in an isolated evaluation harness.
These privileged methods are strictly inaccessible to both the planning agent and the interface compiler.
By separating runtime execution from benchmarking machinery, TaskVM ensures that its evaluation results reflect genuine interactive competence.


\begin{figure*}[t]
\centering
\begin{subfigure}[b]{0.48\textwidth}
\includegraphics[width=\textwidth]{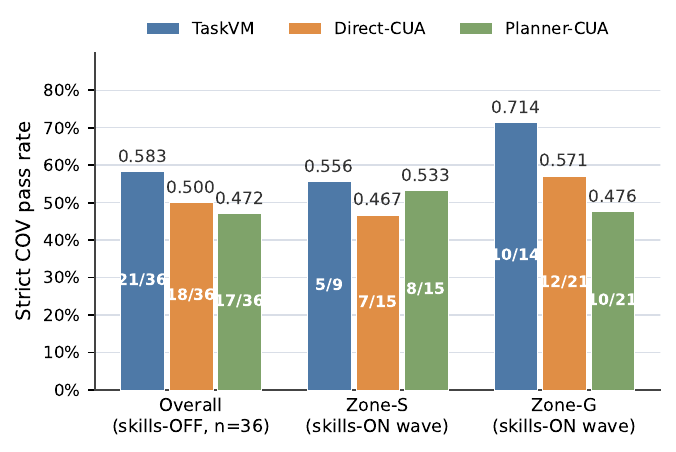}
\caption{Aggregate COV strict pass rate across conditions, overall and by task zone. Per-family outcomes appear in Appendix Figure~\ref{fig:stage1-families}.}
\label{fig:stage1-main}
\end{subfigure}
\hfill
\begin{subfigure}[b]{0.48\textwidth}
\includegraphics[width=\textwidth]{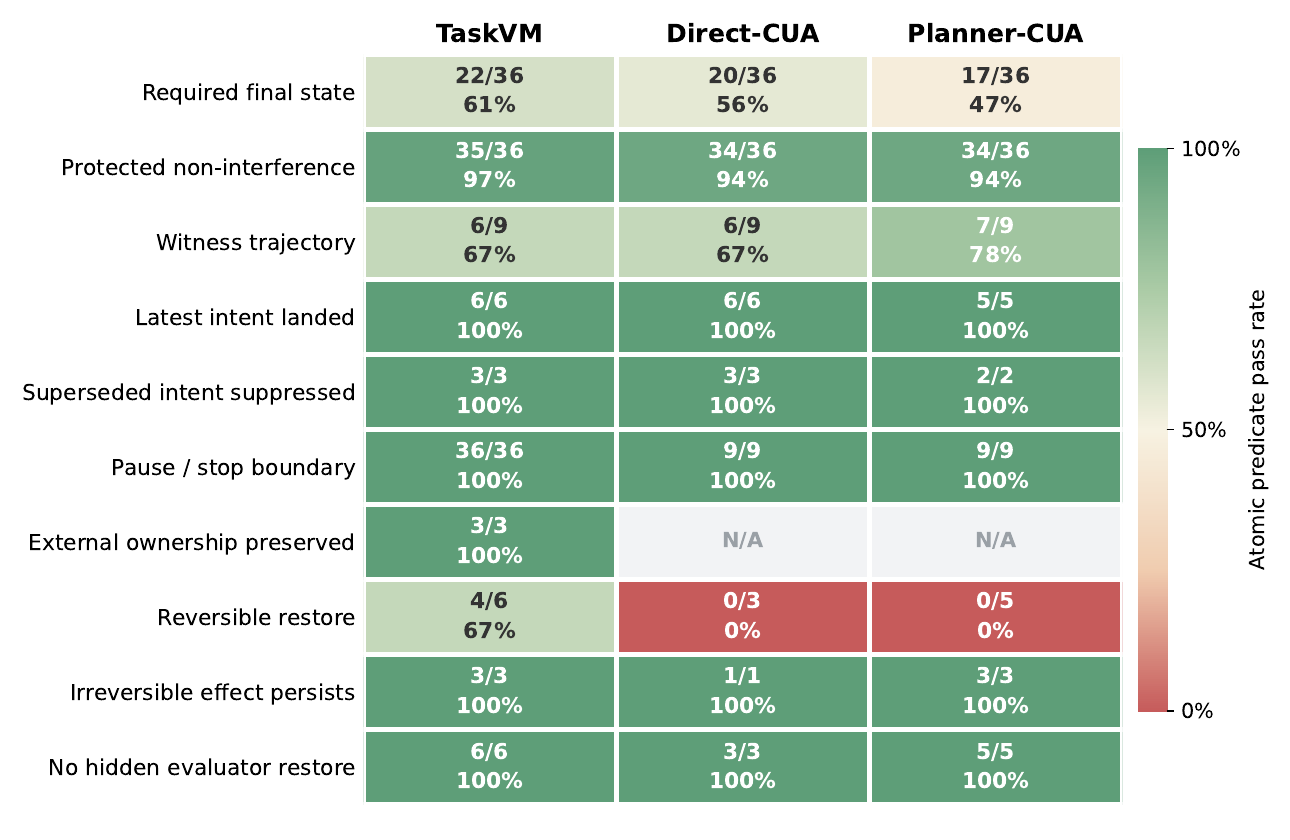}
\caption{Adjudication breakdown across individual COV predicates.}
\label{fig:stage1-heatmap}
\end{subfigure}
\caption{Stage 1 automated technical evaluation results on the frozen suite of 36 task instances spanning 12 structural task families. (a) Overall bars come from the skills-OFF main wave; zone bars come from the skills-ON comparison wave with unmatched denominators, as documented in Table~\ref{tab:zone-split}.}
\label{fig:stage1-results}
\end{figure*}

\section{Stage 1: Automated Technical Evaluation}
\label{sec:stage1}

We evaluate the technical capabilities, operational bounds, and governance properties of TaskVM through an automated benchmark on a frozen suite of multi-application workflows.
This evaluation establishes whether explicit state contracts improve execution reliability over procedural computer-use agents, how performance distributes across task zones, and whether skills injection preserves execution non-inferiority.


\begin{table}[t]
\caption{Stage 1 automated bench under skills injection OFF, 36 task instances per condition. TaskVM achieves the highest Contract-Outcome Verification (COV) strict pass rate and the highest atomic predicate mean; the lower block reports its advantage over each baseline.}
\label{tab:main-comparison}
\scriptsize
\setlength{\tabcolsep}{3pt}%
\begin{tabular}{l c c c c}
\toprule
\textbf{Condition} & \textbf{COV strict rate} & \textbf{COV atomic mean} & \textbf{Model calls} & \textbf{Wall time (s)} \\
\midrule
TaskVM (ours)       & \textbf{0.583} (21/36) & \textbf{0.870} & 1{,}126 & 19{,}960 \\
Direct-CUA          & 0.500 (18/36) & 0.761 & 1{,}361 & 9{,}010 \\
Planner-CUA         & 0.472 (17/36) & 0.737 & 1{,}031 & 6{,}695 \\
Template Control$^\dagger$ & 0.083 (3/36)  & 0.712 & 144     & 8{,}527 \\
\midrule
\multicolumn{5}{@{}l}{\textit{TaskVM advantage over baselines}} \\
\quad vs Direct-CUA  & +0.083 (+16.6\%) & +0.109 (+14.3\%) & $\cdot$ & $\cdot$ \\
\quad vs Planner-CUA & +0.111 (+23.5\%) & +0.133 (+18.0\%) & $\cdot$ & $\cdot$ \\
\bottomrule
\end{tabular}
\\[2pt]
{\footnotesize $^\dagger$ Template Control is an assembly sanity-check driver that replays scripted gestures with zero LLM invocations (the 144 figure above counts scripted action events, not model calls); 101 of its 108 runs across the evaluated waves failed at the architect stub due to an implementation limitation in parsing Chinese natural-language goals, so it is \emph{not} a valid baseline.}
\end{table}

\begin{table}[t]
\caption{COV results split by task zone (skills injection ON). Zone-G families exercise intent landing, supersession, pause boundaries, and reversible restore. Subsets and denominators differ across conditions; rates describe the tested populations and are not matched contrasts.}
\label{tab:zone-split}
\small
\setlength{\tabcolsep}{4pt}%
\begin{tabular}{l c c c}
\toprule
\textbf{Zone / Condition} & \textbf{TaskVM} & \textbf{Direct-CUA} & \textbf{Planner-CUA} \\
\midrule
\multicolumn{4}{l}{\textit{Zone-S: sequential / simple families (\texttt{sqv}, \texttt{fav}, \texttt{lpv})}} \\
\quad COV strict rate   & 0.556 (5/9)  & 0.467 (7/15) & 0.533 (8/15) \\
\quad COV atomic mean    & 0.852        & 0.733        & 0.767        \\
\midrule
\multicolumn{4}{l}{\textit{Zone-G: governance-dense families}} \\
\quad COV strict rate   & \textbf{0.714 (10/14)} & 0.571 (12/21) & 0.476 (10/21) \\
\quad COV atomic mean    & \textbf{0.922} & 0.822 & 0.797 \\
\bottomrule
\end{tabular}
\end{table}

\begin{table}[t]
\caption{Effects of procedural skills injection. Upper block: paired family-level outcome flips between the skills-OFF and skills-ON waves. Lower block: contract-faithful execution (CFE) non-inferiority of TaskVM against its strongest baseline within the skills-ON wave, cluster-bootstrapped (10,000 resamples, 35 paired clusters).}
\label{tab:skills-injection}
\scriptsize
\setlength{\tabcolsep}{3pt}%
\begin{tabular}{l c c c}
\toprule
\textbf{Paired family flip} & \textbf{TaskVM} & \textbf{Direct-CUA} & \textbf{Planner-CUA} \\
\midrule
fail $\rightarrow$ pass (skills helped) & 4  & 0 & 0 \\
pass $\rightarrow$ fail (skills hurt)    & 3  & 0 & 0 \\
unchanged                               & 14 & 18 & 18 \\
missing (no OFF data)                   & 0  & 3  & 3 \\
\midrule
\textbf{Net gain}                       & \textbf{+1} & \textbf{0} & \textbf{0} \\
\midrule
\multicolumn{4}{@{}l}{\textit{CFE non-inferiority, TaskVM vs Direct-CUA (skills-ON wave)}} \\
\quad Point estimate  & \multicolumn{3}{c}{+0.0571} \\
\quad One-sided 95\% lower bound & \multicolumn{3}{c}{$-0.0857$} \\
\quad Verdict at protocol-specified 10pp margin & \multicolumn{3}{c}{Non-inferior} \\
\bottomrule
\end{tabular}
\end{table}

\subsection{Evaluation Protocol and Contract-Outcome Verification}

The automated evaluation suite comprises 36 task instances spanning 12 structural task families derived from the TaskVM architectural contract, with three variants per family.
Execution correctness is adjudicated through \textbf{Contract-Outcome Verification (COV)}, structured across ten machine-verifiable domain predicates defined in Appendix~\ref{app:predicates}.
Each trial is evaluated against all applicable predicates.
These include certifying that target fields reach their required final values (\texttt{required\_\allowbreak final\_\allowbreak state}), confirming that protected fields outside the authorized set remain untouched (\texttt{protected\_\allowbreak non\_\allowbreak interference}), ensuring that executions visit required witness checkpoints (\texttt{required\_\allowbreak witness\_\allowbreak trajectory}), and verifying that superseded user intents leave no residual effects (\texttt{superseded\_\allowbreak intent\_\allowbreak suppressed}).
Furthermore, transactional boundaries are checked by verifying that pause and stop commands halt execution immediately (\texttt{pause\_\allowbreak stop\_\allowbreak boundary}), externally owned objects are not overwritten (\texttt{external\_\allowbreak ownership\_\allowbreak preservation}), and requested rollbacks restore pre-states without hidden evaluator resets (\texttt{no\_\allowbreak hidden\_\allowbreak evaluator\_\allowbreak restore}).
We report two complementary metrics: \textit{strict pass rate}, requiring all applicable predicates in a trial to pass, and \textit{atomic predicate mean}, measuring the average pass rate across applicable predicates.

Trials are evaluated across three primary execution conditions with skills injection disabled, comprising 36 instances per condition: TaskVM, Direct-CUA, and Planner-CUA.
Direct-CUA perceives raw screenshots and dispatches actions directly, receiving interventions as natural language messages delivered to an interactive chat drawer.
Planner-CUA formulates hierarchical natural language plans before dispatching actions.
All conditions execute against identical hidden-oracle world contracts, ensuring that evaluation relies on external ground truth rather than agent self-reports.

\paragraph{Evaluation configuration.}
All LLM roles in all conditions ran the same production model (GPT-5.6-sol) through the same internal inference endpoint, so condition differences reflect the interaction and runtime architecture rather than model choice.
In the TaskVM condition the model serves four prompted roles: state compiler, task architect, CUA executor, and an independent model verifier reading fresh screenshots; Direct-CUA and Planner-CUA use the same model for a single agentic loop (with a planning preamble in the latter) over identical environment seeds.
Template Control is a scripted gesture driver with zero LLM calls.
The user-facing runtime sees only rendered pixels and standard input gestures, whereas oracle-only capabilities (state resets, internal-state reads, event injection) live in an isolated evaluation harness; every trial manifest records the model identity, ledger call counts per role, and wall time, and the frozen run artifacts are retained in the consolidated archive (full prompts and generation settings in the supplementary package).

\subsection{Main Benchmark Findings}

Table~\ref{tab:main-comparison} reports the primary comparison across the 36 task instances, visualized in Figure~\ref{fig:stage1-results}.
On the skills-OFF main comparison, TaskVM obtained 21/36 strict passes (rate: 0.583) and an atomic predicate mean of 0.870, compared with 18/36 (0.500) and 0.761 for Direct-CUA, and 17/36 (0.472) and 0.737 for Planner-CUA.
The advantage amounts to +0.083 (+16.6\%) over Direct-CUA and +0.111 (+23.5\%) over Planner-CUA in strict pass rate, alongside +0.109 (+14.3\%) and +0.133 (+18.0\%) in atomic predicate mean.
Inspection of error traces indicates that baseline failures frequently stemmed from premature completion claims and collateral modifications.
Direct-CUA often emitted stop tokens before asynchronous web updates persisted, and Planner-CUA altered adjacent form fields when re-planning.
In contrast, TaskVM enforced witness predicates that confirmed persisted state before certifying completion.
However, this verification property introduces operational overhead: TaskVM consumed 19,960 seconds of wall time across 1,126 model calls, about 2.2 times the wall time of Direct-CUA, reflecting the latency of round-trip state compilation and verification polling.
We report this machine-side cost as a measured property of the automated harness; it cannot be decomposed here into user-facing costs, which the later studies measure on their own terms.
Table~\ref{tab:main-comparison} also documents Template Control (3/36 strict pass rate), included exclusively as an assembly sanity check; 101 of 108 trials failed at the architect stage due to a limitation in parsing Chinese-language goals, so it is not an empirical baseline.

\subsection{Predicate-Level Disaggregation}

The predicate-level breakdown in Figure~\ref{fig:stage1-heatmap} localizes where the advantage originates.
The largest source of separation is \texttt{required\_\allowbreak final\_\allowbreak state}, which TaskVM satisfied on 22 of 36 instances versus 20 for Direct-CUA and 17 for Planner-CUA.
Two further observations concern recovery specifically.
On \texttt{reversible\_restore}, TaskVM restored the protected pre-state in 4 of 6 applicable trials, while Direct-CUA restored none of its 3 applicable trials and Planner-CUA none of its 5: in the adjudication records the baselines possess no restore verb at all, and on \texttt{external\_\allowbreak ownership\_\allowbreak preservation} they reach no testable scenario.
TaskVM is also the only condition measured on all ten predicates, including 36/36 passes on \texttt{pause\_\allowbreak stop\_\allowbreak boundary}; this full-spectrum measurability is itself a distinguishing property of contract-checked execution.
This machine-side asymmetry, where recovery exists as an executable verb only under TaskVM, is mirrored later at the human layer: recovery operations were the most adopted governance functions in Stage 2 (Section~\ref{sec:stage2}) and the most demanded automation in Stage 3 (Section~\ref{sec:stage3}).

\subsection{Zone Breakdown and Governance Efficacy}

To understand where state contracts provide utility, Table~\ref{tab:zone-split} partitions evaluation across task zones in a separate skills-ON comparison wave.
Zone-S comprises five structural and execution families (Loop \texttt{blv}, Fan-out \texttt{fav}, Partial-failure \texttt{pfv}, Cross-app \texttt{cav}, Sequence \texttt{sqv}), while Zone-G comprises seven governance-dense families (Conflict \texttt{cfv}, UI-drift \texttt{drv}, Interrupt \texttt{itv}, Local-patch \texttt{lpv}, Goal-patch \texttt{gpv}, Irreversible \texttt{irv}, Rollback \texttt{rbv}) that specifically test user intervention intent landing, supersession, pause boundaries, and reversible restore; the same family-to-zone dictionary drives Figure~\ref{fig:stage1-families} and Appendix Table~\ref{tab:s1-instances}.

In Zone-S, TaskVM achieved a strict pass rate of 0.556 (5/9) and an atomic mean of 0.852, compared with 0.467 (7/15) and 0.733 for Direct-CUA, and 0.533 (8/15) and 0.767 for Planner-CUA.
In Zone-G, TaskVM achieved a strict pass rate of 0.714 (10/14) and an atomic mean of 0.922, compared with 0.571 (12/21) and 0.822 for Direct-CUA, and 0.476 (10/21) and 0.797 for Planner-CUA.
On these descriptive rates, the advantage over both baselines is larger in Zone-G than in Zone-S.
The same governance dimension later separates the two interfaces for human users: in Stage 2, participants reported their largest perceived-control gain on the ability to stop and redirect the agent (Section~\ref{sec:stage2}).
Notably, the analyzed subsets and denominators differ across execution conditions (e.g., 9 vs 15 and 14 vs 21).
These unmatched rates are reported to document the tested populations; missing or excluded cases are not treated as zero, and the difference between unmatched rates is not interpreted as a matched statistical contrast.

\subsection{Skills Injection and Non-Inferiority Analysis}

We examine whether procedural skills injection harms TaskVM's contract faithfulness and whether TaskVM remains non-inferior to its strongest baseline once skills are enabled (Table~\ref{tab:skills-injection}).
Paired across 21 instances between the two waves, skills injection flipped four instances in TaskVM's favor and three against it, a modest net gain of +1; neither CUA baseline showed any net change.
Because the skills-OFF and skills-ON waves differ in coverage, we additionally evaluate non-inferiority within the same skills-ON wave on contract-faithful execution (CFE), a per-instance correctness measure derived from the task boundary semantics of each family.
The two analysis collections are explicitly distinct: the strict-COV index retains only fully measured cells (23 of 36 for TaskVM under skills-ON), whereas the CFE comparison additionally admits the protocol's per-family fallback judgments (e.g., boundary arcs judged from pre-boundary oracle segments), yielding 35 comparable instance clusters paired between TaskVM and Direct-CUA; the 35 are instance-level clusters, not a third task set.
A cluster bootstrap over these 35 paired clusters (10,000 resamples, fixed seed) yields a point estimate of the TaskVM minus Direct-CUA difference of +0.0571, with a one-sided 95\% lower bound of $-0.0857$.
This lower bound clears the margin of 10 percentage points specified in our evaluation protocol, establishing non-inferiority at that margin; sensitivity analyses at 5 and 7.5 points remain inconclusive, and the verdict also holds at 15 points.
Among the eight discordant clusters, TaskVM alone succeeded on five instances concentrated in drift-robust rescheduling and sequential contract chains (\texttt{drv-01}, \texttt{drv-03}, \texttt{sqv-01} through \texttt{sqv-03}), while Direct-CUA alone succeeded on three instances involving confusable references and retargeting (\texttt{gpv-03}, \texttt{lpv-03}, \texttt{pfv-03}).
Skills injection therefore does not degrade TaskVM below its strongest baseline on contract-faithful execution, though the inconclusive stricter margins call for larger paired samples.
This result supersedes an earlier cross-wave draft analysis that was inconclusive; the final protocol and its sensitivity margins are documented in Appendix~\ref{app:stage1-honesty}.

\subsection{Stress-Tier Robustness}

Beyond the frozen benchmark waves, a development-era stress tier probed governance durability under adversarial goal mutations with skills enabled throughout.
The tier comprises 18 variants at difficulty levels L3 and L4, covering conflict injection, goal repair, and cross-session intervention, adjudicated under a majority criterion in which a frozen pass verdict is never reversed.
Eleven of 12 L3 variants and all 6 L4 variants passed.
The single L3 failure is \texttt{rbvl3}, a rollback-compensation family blocked by a signing-window race condition that remains an open, documented defect.
We report this tier as robustness evidence rather than held-out results.
The placement of the sole failure is nonetheless informative: rollback is simultaneously the hardest capability to make reliable and, as the next two sections show, the capability human participants used and requested the most (Sections~\ref{sec:stage2} and~\ref{sec:stage3}); we return to this tension in the Discussion.



\begin{table*}[t]
\centering
\caption{Stage 2 within-subject user study evaluation ($N = 37$). We report means, paired differences ($\Delta$), directional participant counts (in favor of TaskVM / in favor of Conversational CUA), Wilcoxon signed-rank $p$-values, and standardized effect sizes ($d_z$). \textbf{Bold} entries are statistically significant at $p < .05$. Write-probe margins are a manipulation check of the replay protocol: the environment and the replayed trajectory were identical across arms, so identical margins indicate that neither arm offered systematically easier intervention opportunities, not that the interfaces would execute with equivalent competence outside the rig.}
\label{tab:user-study}
\footnotesize
\begin{tabular}{@{}l r r r c c c@{}}
\toprule
\textbf{Measure} & \textbf{TaskVM} & \textbf{Conv CUA} & \textbf{$\Delta$} & \textbf{Paired counts} & \textbf{$p$} & \textbf{$d_z$} \\
\midrule
\multicolumn{7}{@{}l}{\textit{Subjective measures (1--7 unless noted; higher is better except NASA-TLX)}} \\
Trust (2-item TIA composite $\uparrow$)          & 5.54 & 4.36 & \textbf{+1.18} & 21 / 7  & \textbf{.0030} & +0.55 \\
Recovery control (4 study-specific items $\uparrow$)& 5.56 & 4.63 & \textbf{+0.93} & 22 / 13 & \textbf{.0050} & +0.51 \\
\quad Stop and redirect (\texttt{rc\_stop\_redirect}) & 5.68 & 4.43 & \textbf{+1.24} & 23 / 7  & \textbf{.0035} & +0.58 \\
\quad Correct errors (\texttt{rc\_correct}) & 5.49 & 4.59 & \textbf{+0.89} & 20 / 10 & \textbf{.0160} & +0.42 \\
\quad Overall control (\texttt{rc\_overall})& 5.35 & 4.35 & \textbf{+1.00} & 19 / 9  & \textbf{.0099} & +0.44 \\
\quad Influence agent (\texttt{rc\_influence})& 5.73 & 5.14 & \textbf{+0.59} & 14 / 8 & \textbf{.0475} & +0.35 \\
Support satisfaction (adapted 3-item ASQ $\uparrow$) & 5.66 & 4.95 & \textbf{+0.70} & 20 / 11 & \textbf{.0222} & +0.42 \\
Perceived transparency (3-item TIA subset$^{\S}$ $\uparrow$)& 5.07 & 4.50 & +0.57 & 20 / 14 & .0809 & +0.30 \\
Usability (UMUX-Lite, rescaled 0--100 $\uparrow$)    & 70.72 & 60.14 & +10.58 & 20 / 12 & .0881 & +0.33 \\
Workload (NASA-TLX, unweighted 6-dim mean $\downarrow$)& 29.82 & 39.08 & $-$9.26 & 21 lower & .0938 & $-$0.36 \\
\quad Physical demand (0--100 $\downarrow$)  & 17.97 & 30.95 & {\boldmath$-$12.97} & 17 lower & \textbf{.0085} & $-$0.50 \\
\quad Frustration (0--100 $\downarrow$)      & 26.49 & 36.89 & $-$10.41 & 19 lower & .0915 & $-$0.32 \\
\midrule
\multicolumn{7}{@{}l}{\textit{Objective behavioral metrics and descriptive calibration}} \\
Overconfidence gap (confidence $-$ accuracy)$^\ddagger$ & +13.2 pp & +25.8 pp & $-$12.6 pp & $\cdot$ & $\cdot$ & $\cdot$ \\
Read-probe question accuracy                & 84.8\% (56/66) & 73.3\% (44/60) & +11.5 pp & 3 / 2 (tie 3) & 1.000 & +0.10 \\
Write-probe boundary pass$^\dagger$          & 34.3\% (12/35) & 34.3\% (12/35) & 0.0 pp & 12 / 12 / 10 / 0 & 1.000 & 0.00 \\
Write-probe end pass$^\dagger$               & 51.4\% (18/35) & 51.4\% (18/35) & 0.0 pp & 18 / 18 / 1 / 0 & 1.000 & 0.00 \\
Median intervention latency (s)             & 85.3 & 65.7 & +19.6 & 21 slower & .4372 & +0.23 \\
Rollback (reset) operations                 & 17 events & 4 events & +13 & 11 / 2 users & $\cdot$ & $\cdot$ \\
\bottomrule
\end{tabular}
\\[2pt]
{\footnotesize $^\dagger$ In the paired-count column, the two write-probe rows report the four-cell split in the order TaskVM-only pass / Conv-only pass / both-fail / both-pass. Boundary analysis keeps 34 pairs, excluding P34 (unscored in both arms), P19 (no TaskVM-arm probe), and P04 (no conversational-arm probe); marginal denominators are 35 per arm because the TaskVM-arm marginal drops P34 and P19, while the conversational-arm marginal drops P34 and P04. End analysis keeps all 37 pairs, counting the two unscored probes as not-passed, so the paired end rate is 18/37 (48.6\%) while the scored marginal rate is 18/35 (51.4\%). McNemar $p = 1.0$ at both time points.}
\\[2pt]
{\footnotesize $^\ddagger$ The gap uses all-correct accuracy among responses carrying a confidence rating: 12/21 (57.1\%) under TaskVM and 9/20 (45.0\%) under the baseline. The read-probe row above counts all answered probes, where TaskVM all-correct accuracy is 13/22 (59.1\%); the one TaskVM response lacking a confidence rating explains the denominator difference.}
\\[2pt]
{\footnotesize $^{\S}$ Mean of three predictability items with the unpredictability item reverse-scored (8 minus the raw rating).}
\end{table*}

\section{Stage 2: User Study on Task Governance}
\label{sec:stage2}

While automated technical evaluations establish system capabilities across benchmark suites, evaluating human governance requires empirical observation of human operators.
We conducted a controlled, within-subject laboratory user study with 37 participants to evaluate how human operators understand, steer, and verify multi-application agent workflows.
The study investigates how compiling live application state into an editable governance surface influences perceived recovery control, trust, and the alignment between confidence and understanding compared to conversational agent interfaces.

\subsection{Study Goals and Hypotheses}

The primary objective of the user study is to evaluate the interactive efficacy and cognitive impact of task-state governance during collaborative execution.
We formulate four research questions:
\begin{itemize}
\item \textbf{RQ1 (State Understanding and Confidence Gap):} Does task-state governance descriptively narrow the gap between user confidence and objective understanding of application state?
\item \textbf{RQ2 (Perceived Recovery Control and Adoption):} Does exposing transactional checkpoints and reversibility locks improve perceived recovery control and encourage adoption of recovery operations?
\item \textbf{RQ3 (Interaction Cost under Execution Control):} When the environment and the agent's replayed trajectory are held constant across conditions, what interaction latency and workload trade-offs emerge from state-level governance?
\item \textbf{RQ4 (Trust and Subjective Satisfaction):} Does the task-state governance console foster higher subjective trust and satisfaction compared to conversational supervision?
\end{itemize}

We hypothesized that TaskVM would significantly increase perceived recovery control and reported trust while narrowing the observed overconfidence gap.
Furthermore, we anticipated that governing state would introduce an observable interaction latency trade-off even with the environment and the replayed trajectory held identical by design.

\subsection{Participants and Ethics}

We recruited 37 participants through a public research-volunteer group maintained by our laboratory; membership in the group is open to anyone, although in practice it skews toward university students and researchers (Appendix~\ref{app:demographics}).
19 participants were assigned to the TaskVM-first order and 18 to the Conversational-first order.
All participants completed both conditions and enter the analysis as a within-subject sample.
Self-reported demographics and agent-related background, collected post-study through voluntary channels, are detailed in Appendix~\ref{app:demographics}.
Of the 37 completers, 36 disclosed gender, age, and affiliation (15 female, 21 male; ages 18--42, $M{=}24.7$) and 35 completed the background questionnaire.
The sample was technically literate but predominantly novice with respect to GUI agents: only 2/35 reported direct computer-use/GUI-agent professional background, and 14/35 had never used a GUI agent or had tried one only once or twice.

This study was conducted in accordance with the ethical guidelines of the local ethics committee. Prior to participation, all subjects provided informed consent, and their confidentiality and anonymity were strictly maintained throughout the study.

Participants who completed the study with valid retained data received compensation of 50 RMB.
Participation was voluntary, withdrawal was possible at any point without penalty, and compensation was not tied to task performance or questionnaire responses.

\subsection{Within-Subject Experimental Design and Apparatus}

The study employed a counterbalanced within-subject design.
Each participant completed multi-application workflows under two conditions, illustrated in Appendix Figures~\ref{fig:interface-conv} and~\ref{fig:interface-taskvm}:
\begin{itemize}
\item \textbf{Conversational CUA (Baseline):} Participants supervise an agent through an application screencast accompanied by an interactive chat drawer, receiving conversational progress updates and issuing natural language steering directives.
\item \textbf{TaskVM (Ours):} Participants govern execution through the governance cockpit (Figure~\ref{fig:interface}): the editable task-state zone displays desired, preserve, and observed fields beside the live application, the progress zone exposes role status and verification reports, and the global governance zone provides lifecycle controls, checkpoints, and rollback.
\end{itemize}

Participants performed eight representative task scenarios spanning collaborative office productivity, event scheduling, customer support record updates, and social communication.
Each scenario encompassed three distinct phases: an initial autonomous execution phase across multiple applications, an injected environmental perturbation such as dynamic UI drift or an unexpected record conflict introduced by a simulated collaborator, and a required user intervention prompting selective modification of target fields while preserving existing records.
Task order and condition assignments were counterbalanced across participants (19 TaskVM-first, 18 Conversational-first) using Latin square permutations.

\subsection{Procedure}

Each condition began with its own onboarding sequence containing an unscored training task (\texttt{r8-train}) so that participants practiced both interfaces to the same degree, followed by a six-minute think-aloud free-exploration period.
In the formal battery each participant completed three of the eight tasks per condition (Table~\ref{tab:s2-battery}); task--condition--order assignments followed the Latin-square schedule, and read/write probe slots were allocated deterministically per task slot in a balanced rotation, so probe exposure was independent of performance.
The condition-end questionnaire (all subjective scales described below) was administered once at the end of each condition; there were no per-task questionnaires.
The experimenter intervened only at scripted perturbation points and through a fallback prompt (e.g., asking the participant to verify the selected item when the scripted model error did not occur), applied identically in both conditions; a logged help channel gave direct guidance on request, and such help was recorded as help usage.
Participant inputs that did not match a scripted replay branch were steered back onto the canonical path by the operator, and these mismatch and resume events are annotated in the session records.
Sessions were run individually.

To isolate the effect of interaction representations on human perception and behavior, the study executed on a high-fidelity Wizard-of-Oz replay rig.
Underlying application screens and event streams were pre-recorded from authentic runs on a simulated application suite, with every frame undergoing automated semantic verification to ensure visible text, status bar timestamps, and pixel evidence aligned with the experimental script.
When participants intervened, the platform matched the input against a canonical script of branching paths, substituting the runtime engine without altering the study runner (\texttt{ConditionRuntime}, \texttt{StudyRunner}).

The replay protocol fixes three things identically across arms: the environment's event sequence, the agent's recorded steps, and the timing and manifestation of errors.
It deliberately fixes nothing about the participant's own intervention: the write-probe score evaluates what the participant did at the intervention point, so paired outcomes legitimately diverge across conditions.
Consistent with this design, marginal pass rates are identical (34.3\%, 12 of 35 scored probes per arm at the boundary; 51.4\%, 18 of 35 at trial end) while the paired record is fully discordant (Section~\ref{sec:stage2-results}): every pass was earned by different participants in the two arms.
We therefore read the identical margins strictly as evidence that neither arm offered systematically easier intervention opportunities, not as evidence that participants experienced identical outcomes, and not as evidence that the two interfaces would execute with equivalent competence outside the rig.

\subsection{Measures and Analysis}
\label{sec:measures}

Subjective scales were collected once per condition through the condition-end instrument (verbatim item texts, sources, and composite rules in Table~\ref{tab:instruments}).
The NASA Task Load Index (NASA-TLX) was administered as six 0--100 sliders and scored as the unweighted mean (RTLX); the performance slider runs from \emph{Perfect} (0) to \emph{Failure} (100), so all six dimensions share the same demand direction with no post-hoc inversion.
UMUX-Lite, the two-item short form of the Usability Metric for User Experience, is rescaled to 0--100 by $\mathrm{mean}(\mathrm{items}-1)/6 \times 100$.
Recovery control is a study-specific composite of four items (influence, stop/redirect, correct, overall).
Trust is a two-item composite (trust, rely) adapted from the Trust in Automation instrument; perceived transparency is a three-item predictability subset of the same instrument with one reverse-scored item.
Support satisfaction adapts the three After-Scenario Questionnaire items to the condition block.
Read probes (three questions each on committed effects, effective goals, and governance facts) were embedded in drift-triggered trials and preceded by a single confidence rating (0--100) for the whole probe; write probes score the participant's own intervention at the scripting boundary and at trial end against the frozen expectation.

All inferential analyses use the participant as the unit: two-sided paired Wilcoxon signed-rank tests for scalar measures and exact McNemar tests for paired binary outcomes (scipy, version pinned in the analysis environment).
Paired differences are summarized by $d_z = \mathrm{mean}(d_i)/\mathrm{sd}(d_i)$; reported confidence intervals are $t$-based intervals of the mean paired difference.
Missing responses per measure are reported in the table (valid pairs range from 34 to 37 for write probes and 8 for read-probe pairing); zero differences are handled by the implementation's defaults.
We report the complete set of planned comparisons without selective omission, with unadjusted $p$-values interpreted alongside effect sizes, and we do not retroactively designate a subset as the sole primary endpoints.
All numbers are reproducible from the consolidated archive's analysis scripts.

\subsection{User Study Results}
\label{sec:stage2-results}

Table~\ref{tab:user-study} summarizes the primary subjective psychometric scales and objective behavioral metrics across the 37 participants.
Analysis rules follow Section~\ref{sec:measures}.

Participants reported significantly higher perceived recovery control under TaskVM than under the conversational baseline.
On a composite of four study-specific recovery-control items (1 to 7), TaskVM achieved a mean rating of 5.56 compared to 4.63 for the conversational baseline ($\Delta = +0.93$, Wilcoxon $p = .0050$, $d_z = +0.51$, with 22 participants favoring TaskVM versus 13 favoring Conversational CUA).
All four constituent items demonstrated statistically significant advantages for TaskVM: ability to stop and redirect the agent (\texttt{rc\_stop}: 5.68 vs 4.43, $\Delta = +1.24$, $p = .0035$, $d_z = +0.58$), general recovery control (\texttt{rc\_overall}: 5.35 vs 4.35, $\Delta = +1.00$, $p = .0099$, $d_z = +0.44$), ability to correct mistakes (\texttt{rc\_correct}: 5.49 vs 4.59, $\Delta = +0.89$, $p = .0160$, $d_z = +0.42$), and perceived influence over agent actions (\texttt{rc\_influence}: 5.73 vs 5.14, $\Delta = +0.59$, $p = .0475$, $d_z = +0.35$).
This perceived empowerment was accompanied by concrete control use.
Goal interventions (pause, edit, resume) were exercised by design in every task.
Beyond them, 11 of 37 participants issued at least one rollback (reset) in the TaskVM condition, totaling 17 rollback events against 2 participants and 4 events in the conversational baseline; 13 of the 17 TaskVM events, issued by 8 of the 11 participants, occurred outside the two tasks that scripted an undo request (Table~\ref{tab:s2-battery}).
This asymmetric adoption mirrors the machine-side asymmetry established in Section~\ref{sec:stage1}, where recovery existed as an executable verb only under TaskVM; what the benchmark showed baselines cannot do became, for participants, the capability they reached for first.
We emphasize that this reflects perceived recovery control and control use; the replayed trajectory the participants watched was identical across conditions, so these differences cannot be attributed to what the two arms showed them.

Objective intervention outcomes corroborate this framing as a manipulation check rather than a treatment effect (Table~\ref{tab:user-study}).
At the intervention boundary, 12 participants passed only under TaskVM and 12 only under the baseline, 10 failed under both, and 0 passed under both ($n = 34$ pairs with both arms scored); at trial end the corresponding split is 18, 18, 1, and 0 across all 37 pairs.
Every pass was therefore discordant under trajectories the rig held identical, indicating that whether an intervention landed tracked individual intervention behavior rather than arm-level execution differences.

Subjective trust and post-task satisfaction were significantly higher under TaskVM.
On a two-item trust composite (trust and reliance items) adapted from the Trust in Automated Systems instrument, TaskVM attained a mean rating of 5.54 versus 4.36 for the conversational baseline ($\Delta = +1.18$, $p = .0030$, $d_z = +0.55$, with 21 participants rating TaskVM higher versus 7 rating Conversational CUA higher).
Support satisfaction measured through an adapted three-item After-Scenario Questionnaire covering ease, completion time, and informational support was likewise significantly higher (5.66 vs 4.95, $\Delta = +0.70$, $p = .0222$, $d_z = +0.42$).
Perceived transparency, computed from a three-item predictability subset with one reverse-scored item, was directionally higher (5.07 vs 4.50, $\Delta = +0.57$, $p = .0809$, $d_z = +0.30$), and usability on UMUX-Lite, rescaled to a 0 to 100 range through the standard two-item transformation, followed the same directional pattern without reaching significance (70.72 vs 60.14, $\Delta = +10.58$, $p = .0881$, $d_z = +0.33$).

Task-state governance descriptively reduced the observed confidence-accuracy gap.
During drift trials where objective read probes were administered, participants reported subjective confidence ratings before submitting answers.
Under Conversational CUA, participants reported a mean confidence of 70.8\% while achieving an actual all-correct accuracy of 45.0\%, producing an overconfidence gap of +25.8 percentage points.
Under TaskVM, participants reported a comparable mean confidence of 70.4\% but achieved an actual accuracy of 57.1\%, narrowing the overconfidence gap to +13.2 percentage points.\footnote{All-correct accuracies used in the gap exclude responses missing the confidence rating: 12/21 (57.1\%) under TaskVM and 9/20 (45.0\%) under the baseline. Counting all answered probes yields 13/22 (59.1\%) for TaskVM; the single TaskVM response lacking a confidence rating accounts for the denominator difference.}
On individual probe questions, TaskVM achieved an accuracy of 84.8\% (56/66) versus 73.3\% (44/60) for Conversational CUA across committed effects, effective goals, and governance facts.
However, because drift events triggered probes on a subset of trials, matched participant pairs were limited ($n = 8$, Wilcoxon $p = 1.0$, $d_z = +0.10$).
Consequently, this probe accuracy stands as directional evidence rather than a statistically confirmed gain in state understanding, and the narrower gap represents a descriptive reduction in overconfidence rather than established calibration.
Furthermore, participants in the conversational condition left twice as many probes unanswered (6 vs 3), frequently chatting past probe triggers without inspecting underlying state.

Interacting with state controls introduced an observed latency trade-off.
Median intervention latency was 85.3 seconds under TaskVM compared to 65.7 seconds under Conversational CUA ($\Delta = +19.6$ seconds in the mean paired difference, 95\% $t$-based CI of that mean difference [$-9.2$, $+48.5$], Wilcoxon $p = .4372$, $d_z = +0.23$).
This observed difference showed substantial variation across participants (21 slower, 16 faster) and was not statistically reliable.
We treat this observed pattern as an interaction trade-off rather than an established cognitive mechanism: interacting with structured state projections naturally involves reviewing visual representations before committing edits, whereas conversational chat permits immediate text dispatch.

Subjective workload on the NASA-TLX exhibited a directional reduction under TaskVM.
Total workload, computed as the unweighted mean of the six dimension ratings, was lower for TaskVM (29.82 vs 39.08, $\Delta = -9.26$, Wilcoxon $p = .0938$, $d_z = -0.36$, with 21 participants reporting lower workload).
All six NASA-TLX subscales showed directionally lower demand for TaskVM, with physical demand reaching statistical significance (17.97 vs 30.95, $\Delta = -12.97$, $p = .0085$, $d_z = -0.50$) and frustration trending lower (26.49 vs 36.89, $\Delta = -10.41$, $p = .0915$, $d_z = -0.32$).
The largest perceived-control gain, the ability to stop and redirect the agent, aligns with the governance-dense zone where TaskVM's machine-side advantage concentrated in Section~\ref{sec:stage1}, suggesting that automated adjudication and human perception are sensitive to the same governance dimension.
Mental demand (41.49 vs 49.73, $p = .2507$), temporal demand (23.24 vs 31.76, $p = .1121$), and effort (38.11 vs 44.46, $p = .2489$) remained directionally consistent.
Analysis of order sequence showed no reversal of the primary pattern across order groups: participants beginning with TaskVM reported a mean workload difference of $\Delta\text{TLX} = -11.5$, while those beginning with Conversational CUA reported $\Delta\text{TLX} = -6.9$, with parallel directional consistency across UMUX-Lite and TIA trust scores.

Stage 2 held the environment and the replayed agent trajectory constant by design, which isolates perception but leaves open whether these gains persist when a live model, rather than a replay rig, executes tasks the participants themselves invent.
Stage 3 (Section~\ref{sec:stage3}) examines exactly this open-ended setting.


\begin{figure*}[t]
\centering
\includegraphics[width=\textwidth, trim=26 205 25 198, clip]{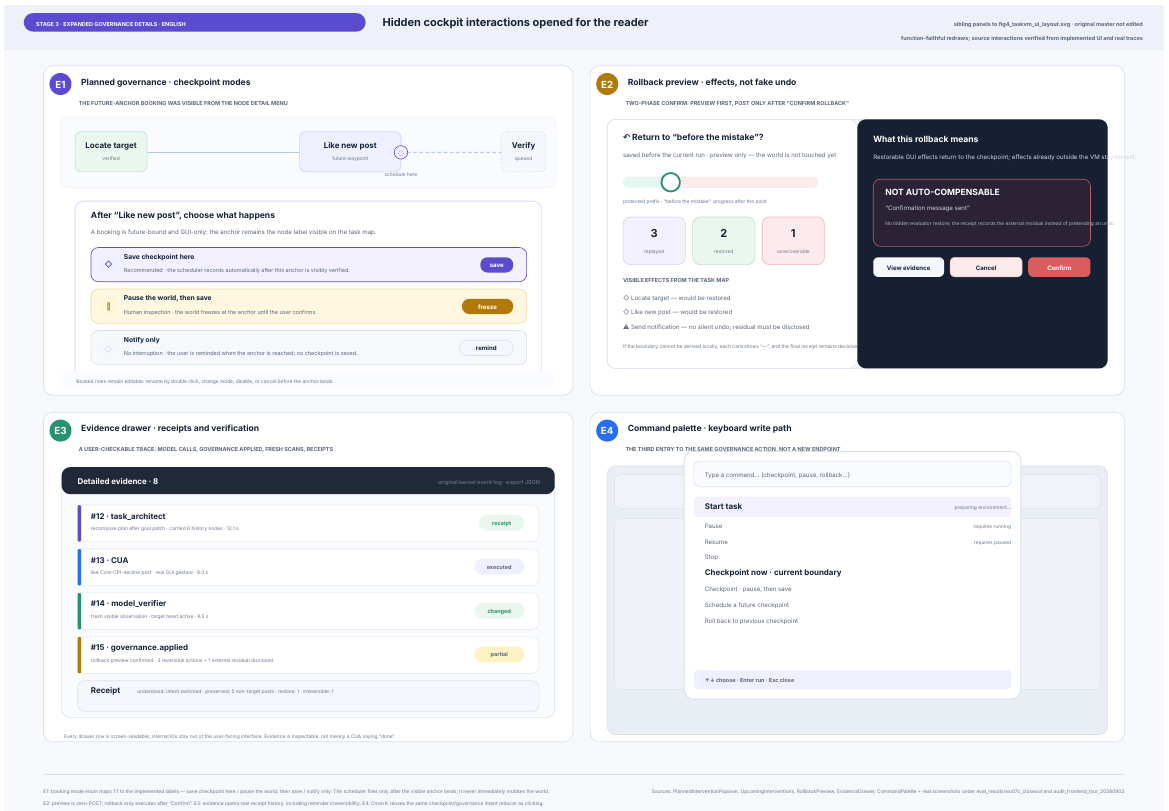}
\caption{Expanded governance interactions in TaskVM, shown as explanatory English redraws. (E1) Planned checkpoint modes. (E2) A rollback preview distinguishing restorable effects from non-compensatable residuals. (E3) An evidence drawer presenting execution and verification records. (E4) A command palette for governance actions. These panels illustrate interface functions; they are not a quantitative summary of Stage 3 use.}
\label{fig:stage3-expanded}
\end{figure*}

\section{Stage 3: Open Exploration with a Live Model}
\label{sec:stage3}

While Stage 2 isolates perception under controlled replay, it leaves open how the governance surface is used when a live model executes tasks that users invent themselves.
Stage 3 addresses this question through an exploratory follow-up in which fourteen of the 37 Stage 2 participants returned to author their own tasks and govern live-model execution through the TaskVM cockpit (Figure~\ref{fig:stage3-expanded}), with no conversational control condition.
Stage 2 carries the controlled comparisons of this paper; Stage 3 characterizes use, breakdowns, and requested improvements once the scripting rails were removed.


\begin{table}[t]
\caption{Stage 3 exploratory follow-up with a live model on participant-authored tasks ($N = 14$ returning participants). NASA-TLX total workload is paired against each participant's own Stage 2 scores. All 14 participants scored lower in Stage 3 than in their earlier conversational-arm sessions. Comparisons are reported descriptively, since task selection, practice, study procedures, and execution mode changed together across stages.}
\label{tab:stage3-tlx}
\scriptsize
\setlength{\tabcolsep}{3pt}%
\begin{tabular}{l c c c c}
\toprule
\textbf{Paired contrast} & \textbf{mean $\Delta$} & \textbf{$d_z$} & \textbf{lower / higher} & \textbf{Wilcoxon $p$} \\
\midrule
Stage 3 vs Stage 2 TaskVM arm        & $-9.70$  & $-0.83$ & 12 / 2  & .011 \\
Stage 3 vs Stage 2 Conversational arm & $-34.40$ & $-1.43$ & 14 / 0  & $<$.001 \\
Stage 3 vs Stage 2 two-arm mean       & $-22.05$ & $-1.80$ & 14 / 0  & $<$.001 \\
\bottomrule
\end{tabular}
\end{table}

\subsection{Design of the Open Exploration}

Each returning participant issued one or more tasks of their own choosing, spanning messaging, scheduling, video browsing, and music and content applications on the same simulated application suite used in the earlier stages; per-participant usage volumes appear in Table~\ref{tab:s3-usage}.
Execution was driven by the live production model pipeline rather than a replay rig, so failures, retries, and recovery needs arose naturally.
Participants interacted with the full governance surface, including live state projection, mid-flight goal edits, checkpoints, and rollback.

Two properties of the study environment shaped what participants could attempt, and we disclose both because they delimit the generality of the Stage 3 material.
First, the substrate is a static, offline world: every searchable object exists only because it was injected during setup, so open-ended search tasks could hit coverage gaps.
When a requested object was absent, strict goal binding prevented the runtime from silently substituting a near-match, and the verification layer reported the shortfall rather than fabricating success. Participants experienced this as a failed search, but it is exactly the honest-failure property the harness is designed to guarantee, and it locates the frustration in substrate coverage rather than in the governance layer.
Second, for messaging tasks participants chose recipients from a designated set of fictional contacts, because the injected contact graph is finite.
This constrains only the binding target. Task structure, ordering constraints, and the governance operations exercised are unaffected, and the identity of a message recipient is immaterial both to GUI execution difficulty and to the governance claims under study.
After each session we collected the NASA-TLX workload instrument and two open-ended questions about the experience and desired improvements; no read or write probes were embedded.

Invitation and compensation followed a deliberate sequence that matters for interpretation.
At the end of Stage 2, participants were invited to the optional follow-up with an explicit statement that no additional compensation would be offered; this was a deliberate design choice to minimize payment-motivated self-selection.
Fourteen of 37 participants returned.
Post hoc, we observed that Stage 3 sessions incurred time and effort comparable to Stage 2, in some cases longer, and we therefore compensated all Stage 3 completers at the same rate (50 RMB, in addition to the 50 RMB for Stage 2).
This decision was made after data collection, applied uniformly, and was not contingent on task performance, questionnaire responses, or evaluation outcomes.

The returning subsample was also biased toward benefit.
In Stage 2, returning participants had shown a much larger workload difference between arms (TaskVM minus Conversational mean of $-24.70$ points) than non-returning participants ($+0.14$), meaning that those who came back were precisely the participants who had already experienced the largest TaskVM advantage.
We therefore interpret all Stage 3 outcomes descriptively: they characterize how an already-advantaged, self-selected subsample used the interface on self-authored tasks, and they do not estimate effects for the full Stage 2 population.
That 14 of 37 participants returned under an explicit no-compensation promise is consistent with the interface carrying engagement value, although self-selection makes this reading tentative.

\subsection{Workload Under a Live Model}

Stage 3 workload ratings were lower than the same participants' earlier ratings, and we report this pattern strictly descriptively.
Table~\ref{tab:stage3-tlx} pairs each participant's Stage 3 NASA-TLX total with their own Stage 2 scores.
Stage 3 workload averaged 15.5 on the 0 to 100 scale, against 25.2 for the same participants in their Stage 2 TaskVM arm (mean $\Delta = -9.70$, $d_z = -0.83$, Wilcoxon $p = .011$, with 12 of 14 participants lower) and 49.9 in their Stage 2 Conversational arm (mean $\Delta = -34.40$, $d_z = -1.43$, $p < .001$, with all 14 lower).
Because task selection, practice, study procedures, and execution mode all changed together between stages, these differences cannot be attributed to any single factor, and we do not treat them as evidence of workload reduction.
We report them as low observed workload in these particular sessions, not as evidence about the governance surface's cost in general.
The two participants whose workload rose (P20, P02) attributed it to a model-side execution loop on a complex task and to friction in starting successive tasks, respectively; neither attributed it to governance mechanisms.

\subsection{Open-Ended Responses}

The qualitative material consists of responses to two open-ended questions, not interviews.
We grouped responses into themes, with the individual response as the coding unit; a single participant could contribute to more than one theme, and theme counts below refer to distinct participants.

On the supportive side, four participants (P21, P37, P27, P25) described faithful execution and precise intent recognition under open-ended tasks.
P27 wrote that the system recognized the intended edit and completed the keyword changes without over-guessing what they wanted to do [translated], and P21 described watching the planned decomposition land step by step as reassuring [translated].
One participant's account constitutes logged governance use: P30 inserted a new instruction midway through execution, asking the agent to additionally ask a specific colleague about overtime availability, and reported that the agent completed the original task and then correctly located the colleague and delivered the inquiry, describing the outcome as deeply satisfying [translated]; this insertion also appears in the session record, so it reflects logged use rather than praise alone.
Separately, P35 praised the topology view of task structure as something they really liked [translated]; we count this as appreciation of the interface, not as use of a governance operation.
Two participants (P30, P27) independently requested stronger governance automation, asking that the system place checkpoints automatically at nodes where failures are likely [translated].
We read these requests as demand-side evidence that the checkpoint concept had been absorbed and extended by users, not as reports of a system defect.

On the critical side, three themes emerged, ordered by severity.
First, three participants (P35, P06, P13) were frustrated when open-ended search queries missed the simulated platform's injected data.
One searched for several popular videos and found none of them [translated]; another searched for their own uploaded videos and felt discouraged when nothing came back [translated].
This reflects the finite data coverage of the static offline substrate (Section~\ref{sec:stage3}), not the governance mechanism: in each case the system reported the miss honestly instead of substituting a plausible-looking result, and it is the most visible limitation of the study environment.
Second, four participants (P10, P20, P13, P22) traced failures to live-model execution problems such as retry loops, repeated actions, and lost task continuation.
These failures originate in the executor (model) layer and would be inherited by any interface paradigm; what differs across interfaces is how such failures are surfaced, contained, and recovered from, which is precisely where the governance features are exercised.
We therefore report them as executor-side occurrences with interface-side consequences, without claiming that the governance layer is unaffected by them.
Third, seven participants reported one-iteration usability issues such as progress visibility during retries, state-clearing feedback, and onboarding gaps, and two of them asked for features that already existed but had not been covered in the tutorial.

Across all open-ended responses, no participant spontaneously reported difficulty understanding the state of the underlying applications or losing control of the running task; every negative remark concerned substrate data coverage, executor performance, or interface polish.

\subsection{Cross-Substrate Capability and Boundaries}

A central objective of TaskVM is substrate independence: ensuring that task-state governance operates across diverse software stacks without requiring application-specific programming interfaces.
As mapped in the substrate port of Figure~\ref{fig:arch}, TaskVM abstracts external software through a unified port interface that renders pixels into observations and dispatches standard hardware inputs, and we maintain adapters for modern single-page web applications, mobile environments hosted through MobileGym \citep{wu2026mobilegym}, and Linux desktop applications within OSWorld \citep{xie2024osworld}.
Across these environments, the runtime compiles the same state-level intents into substrate-specific gestures, and the independent visual verifier confirms terminal states from raw screen pixels, so a unified task contract survives platform changes.

Our cross-substrate investigations also expose clear operational boundaries where the abstraction encounters friction:
\begin{itemize}
\item \textbf{Canvas-Rendered Interfaces:} Applications rendered entirely through low-level graphical canvases lack discrete visual text and standard widget affordances, challenging the visual entity extraction pipeline.
\item \textbf{Transient Dynamic Popups:} Short-lived visual notifications that appear and fade within sub-second intervals can elude periodic screenshot polling, occasionally delaying conflict detection.
\item \textbf{Defensive Anti-Automation Gates:} Web platforms deploying CAPTCHA challenges, biometric verification, or aggressive bot-detection scripts intentionally resist automated inputs, representing an absolute boundary requiring direct human handoff.
\item \textbf{Desktop Window Occlusion:} Overlapping application windows on multi-tasking desktops can temporarily obscure visual evidence, requiring the runtime to synthesize window-focus management gestures before polling visual artifacts.
\end{itemize}
Documenting these boundaries clarifies that task-state governance is designed for structured productivity workflows, establishing transparent expectations for open-world deployment.


\section{Discussion}

The core question raised by TaskVM is not merely how agents execute software, but how users interact with, verify, and recover from software state changes with bounded cognitive effort.
Here we discuss four critical interaction design choices emerging from the TaskVM paradigm, its technical evaluations, and its associated operational costs.

\subsection{Defining Preservation Constraints versus Monitoring Actions}

In traditional conversational agents, users are placed in an unnatural supervisory role: watching an automated cursor move across windows and reading lengthy logs to verify execution.
This introduces passive monitoring costs and opportunities for automation bias \citep{bainbridge1983ironies, parasuraman2010bias}.

TaskVM replaces this monitoring with explicit contract definition.
The user specifies both a desired state and a preservation scope, effectively transferring the effort from watching cursor clicks to writing state constraints.
While this shifts human attention from procedural navigation to declarative intent, it does not entirely remove cognitive effort.
Determining which fields may change and which must remain protected is its own conceptual workload.
We argue that when dealing with multi-application environments, defining bounded state invariants is more tractable than mentally synthesizing low-level execution traces across isolated application silos.

\subsection{The Latency and Cost of Verification}

The properties provided by TaskVM do not come without overhead.
In conventional computer-use agents, execution progresses through an unconstrained forward loop, generating actions until the model emits a stop token.
TaskVM, by contrast, introduces structured overhead at multiple stages: compiling initial state, projecting the interface through a secondary model, filtering internal identifiers, and polling rendered screens.

Our automated benchmarks highlight this trade-off.
TaskVM consumed 19,960 seconds of total wall time and 1,126 model calls across 36 task trials, compared to 9,010 seconds / 1,361 calls for Direct-CUA, and 6,695 seconds / 1,031 calls for Planner-CUA (Table~\ref{tab:main-comparison}).
This cost distribution cannot be attributed solely to verification: state compilation, dynamic re-projection, and structured execution boundaries all contribute to the measured wall-time gap.
However, if an interface treats an external task merely as an asynchronous background job, the cost of an unverified error in high-stakes enterprise software vastly outweighs the latency of checking.
An observed trade-off also emerged in our user study: participants exhibited a median intervention latency 19.6 seconds higher under TaskVM (Table~\ref{tab:user-study}), a pattern characterized by substantial variance across users without a statistically reliable condition effect.
Neither of our human studies decomposes this runtime overhead into machine and human shares.
The replay study estimates interaction latency in a controlled setting and cannot establish whether verification delays surface for users under live execution; the exploratory follow-up offers workload descriptions rather than a controlled cost comparison.
A design implication is therefore to make verification effort and its products visible at the points where users must judge or intervene, while explicitly evaluating how evidence presentation, waiting time, and recovery demands interact under live execution.

\subsection{What Can Visible Observations Certify?}

A recurring theme in TaskVM is that interfaces cannot simply trust self-reported execution successes.
We rely on independent visual verification against observable fields to certify task state.
However, this relies on an important boundary: visible observations can prove what is rendered on the display, but they cannot strictly prove hidden backend states or atomic transactional assurances across third-party platforms.

For example, a verified booking screen shows that the UI reflects a date change, but does not inherently prove that airline system databases successfully processed a refund or cancellation ticket.
TaskVM's live state projection surfaces exactly what the system can observe, flagging when visual evidence is insufficient rather than silently assuming consistency.
This transparency highlights that interface trust should be grounded in explicit model boundaries, not in exaggerated claims of system-wide correctness.

\subsection{Designing for Bounded Recovery and Irreversibility}

A prevalent tendency in automated agent design is the masking of failures through optimistic self-reporting.
When an agent makes an irreversible error, such as deleting a contact record, conversational interfaces often provide vague reassurances, misleading users into believing the action can be undone.
TaskVM implements a tripartite taxonomy of reversibility.
Reversible and compensatable actions are tracked and counteracted through inverse GUI gestures, while irreversible operations are explicitly locked to prevent deceptive undo representations.
Empirically, participants exercised goal interventions in every task by design, and 11 of 37 additionally invoked rollback a total of 17 times under TaskVM (baseline: 2 participants, 4 times), with 13 of those 17 events occurring outside the two tasks that scripted an undo request, demonstrating that participants reached for the exposed recovery control even when not asked to.

We note that compensating actions, such as sending a cancellation message, do not equate to true atomic rollbacks.
They introduce secondary effects and may leave behind specific audit trails in third-party systems.
By forcing the interface to distinguish between these recovery modes, TaskVM shifts user trust away from fictional comprehensive undo capabilities toward calibrated expectations about what can, and cannot, be reversed.

Our evidence also reveals a productive tension: rollback is simultaneously the most demanded governance capability and the hardest to make reliable.
Rollback was the operation our baselines lacked entirely, the function 11 of 37 study participants invoked (17 events) beyond scripted requests, and the automation Stage 3 participants requested in stronger form, asking for checkpoints placed automatically at failure-prone nodes.
Yet the one unresolved failure in our stress tier is precisely a rollback-compensation family, blocked by a race between a signing window and the confirmation gesture.
We read this convergence not as a defect to be excused but as the clearest roadmap the paradigm offers: honest vocabulary about recovery creates demand for reliable recovery, and the engineering agenda should follow the demand.


\section{Limitations and Future Work}
\label{sec:limitations}

Our results are bounded in several ways.
Stage 2 holds the environment and the replayed agent trajectory constant, which isolates perception but cannot evidence live-model execution success or real recovery capability; its read probes reached only a subset of trials ($n = 8$ matched pairs), so probe accuracy is directional evidence rather than an established gain in understanding.
Stage 3 restores a live model and participant-authored tasks but removes the control arm, so its lower workload is descriptive and confounded with task freedom and practice, and its sample is self-selected: the 14 returning participants had shown a substantially larger Stage 2 workload advantage than the 23 who did not return ($-24.70$ versus $+0.14$ NASA-TLX points).
Participants were recruited from a single laboratory research-volunteer pool skewing toward university students and researchers.
All stages ran on an offline simulated application suite with finite injected data, so open-ended search occasionally missed the corpus, adding noise without affecting the governance mechanism; cross-platform deployment is engineering-supported through our substrate adapters but not empirically demonstrated here; live-model execution failures such as retry loops originate in the executor layer, yet their consequences still require an interface that can surface and govern them.
On the system side, our compiled schemas express structured productivity workflows whose state maps to discrete variables and invariant predicates; continuous creative workflows (video editing, 3D modeling, freehand illustration) resist this decomposition, and our purely visual polling can miss sub-second transient UI states, complicating fine-grained causal diagnosis.
Finally, TaskVM deliberately does not circumvent defensive gates such as CAPTCHAs; graceful human handoff at such boundaries, broader and more diverse samples, production substrates, and equivalence-controlled live-model comparisons are the natural next steps.


\section{Conclusion}

As autonomous agents gain the ability to execute complex workflows, the object users must govern is not the agent's procedure but the state of the world the agent touches.
We present task-state governance, an interaction model that frames a cross-application task as an editable relationship among desired outcomes, protected facts, and observed application state.
TaskVM instantiates this relationship through task-level controls, separately checked observations, and explicitly bounded recovery.

Our evaluation characterizes three distinct questions with three distinct designs: technical capability and cost on a fixed benchmark suite, subjective experience under controlled replay, and exploratory use by returning participants on self-authored tasks.
These are complementary forms of evidence rather than three confirmations of one mechanism, and each carries the boundaries registered in the preceding section.
Together they show that the obligations and limits of agent-mediated change can be made explicit objects of interaction, verifiable in principle and inspectable in use, while live-execution effectiveness and broader generalization remain open to further study.
By shifting the unit of interaction from procedural plans to verifiable state contracts, TaskVM suggests how users can govern autonomous agents with clarity and calibrated trust: humans govern tasks, while agents operate applications.

\section*{Data Availability}

Analysis scripts and de-identified aggregate data will be released with the supplementary materials.
Identity mappings and raw screen recordings are withheld to protect participant privacy.

\bibliographystyle{ACM-Reference-Format}
\bibliography{taskvm_verified_references}

\clearpage
\appendix

\section{Contract-Outcome Verification Predicates}
\label{app:predicates}

This appendix defines the ten Contract-Outcome Verification (COV) predicates used to adjudicate every Stage 1 trial.
Each trial is scored on every applicable predicate; a strict pass requires all applicable predicates to pass.

\begin{table}[h]
\caption{COV predicates and what each certifies.}
\label{tab:cov-predicates}
\small
\begin{tabular}{p{0.32\linewidth} p{0.58\linewidth}}
\toprule
\textbf{Predicate} & \textbf{Meaning (what it certifies)} \\
\midrule
\texttt{required\_\allowbreak{}final\_\allowbreak{}state} & The target fields specified by the task end in the desired state. \\
\texttt{protected\_\allowbreak{}non\_\allowbreak{}interference} & No observable field in the protected set was modified. \\
\texttt{required\_\allowbreak{}witness\_\allowbreak{}trajectory} & The execution visited the witness checkpoints that prove the right path was taken. \\
\texttt{latest\_\allowbreak{}intent\_\allowbreak{}landed} & The most recent user intent is the one reflected in the final world. \\
\texttt{superseded\_\allowbreak{}intent\_\allowbreak{}suppressed} & A superseded intent did not leave residue effects behind. \\
\texttt{pause\_\allowbreak{}stop\_\allowbreak{}boundary} & A pause or stop was honored at its transactional boundary. \\
\texttt{external\_\allowbreak{}ownership\_\allowbreak{}preservation} & An externally owned object was not silently overwritten. \\
\texttt{reversible\_\allowbreak{}restore} & A requested rollback restored the protected pre-state. \\
\texttt{irreversible\_\allowbreak{}persisted} & An irreversible action was honestly persisted, not hidden. \\
\texttt{no\_\allowbreak{}hidden\_\allowbreak{}evaluator\_\allowbreak{}restore} & No silent evaluator-only restore masked a failure. \\
\bottomrule
\end{tabular}
\end{table}

\section{Stage 1 Data Quality Notes}
\label{app:stage1-honesty}

We document the exclusion and correction decisions behind the Stage 1 numbers so the reported tables can be audited cell by cell.

Five trials in the first wave failed at the infrastructure level rather than at task execution (two vendor server errors and three setup timeouts); these were re-run cleanly, and their invalidated records were removed from all denominators.
Two further trials misclassified by the fail-loud grading gate (one true pass and one valid failure) were vindicated in a subsequent envelope audit and counted correctly.
Frozen pass verdicts are regression-locked and can never be reversed by later waves.
In the skills-ON wave, TaskVM has valid cells for 23 of 36 instances; the remaining cells are recorded as missing rather than scored as failures, and a cell-level index cross-validation across all 108 cells reported zero mismatches.

The skills-ON wave initially produced a misleading side artifact: a cell-level index in which all baseline cells were marked as failures due to a residual obsolete progress gate.
The authoritative rollup supersedes that index (Direct-CUA 19/36, Planner-CUA 18/36); any aggregation from the raw cells would reproduce the obsolete gate and is excluded by an on-record erratum.

Finally, the non-inferiority analysis reported in Table~\ref{tab:skills-injection} is the corrected protocol version computed after adjudication of the frozen waves.
An earlier draft compared the skills-ON and skills-OFF waves against each other and was inconclusive; the corrected analysis follows the evaluation protocol in comparing TaskVM against its strongest baseline within a single wave on contract-faithful execution, pairing 35 task clusters and resampling 10,000 times with a fixed seed.
At the 10 percentage-point margin specified in our evaluation protocol the one-sided 95 percent lower bound ($-0.0857$) clears the margin, establishing non-inferiority; sensitivity analyses at 5 and 7.5 points remain inconclusive, and the verdict also holds at 15 points.
The 23-of-36 valid strict-COV cells and the 35 CFE clusters are explicitly different analysis collections: the former retains only fully measured cells, whereas the latter additionally admits the protocol's per-family fallback judgments, so the larger CFE count does not contradict the smaller COV count, and the 35 clusters are instance-level pairings between TaskVM and Direct-CUA, not a separate task set.

\section{Comparison with Prior Systems}
\label{app:related-matrix}

This appendix positions TaskVM among representative interactive agent systems along one organizing axis: the primary object each interface asks the user to govern.
For each system we trace the consequences of that choice, namely what a user edit commits the system to, what counts as completion, and what recovery can mean.
These are differences in kind, not rankings: the rows are not benchmarked under a shared protocol, and each cell characterizes the system as presented by its authors; where a column falls outside a system's reported model, we mark it as such rather than implying incapability.

\begin{table*}[t]
\centering
\caption{TaskVM in the design space of interactive agent and malleable interface systems, organized by the primary object the user governs. The remaining columns trace what follows from that object: the obligation an edit creates, the evidence that counts as completion, and the meaning of recovery. The final row states the state-contract stance examined in this paper.}
\label{tab:related-matrix}
\footnotesize
\begin{tabular}{@{}p{1.4cm}p{2.5cm}p{2.5cm}p{2.4cm}p{2.4cm}p{1.6cm}@{}}
\toprule
\textbf{System} & \textbf{Governing object (what a task is)} & \textbf{Obligation created by an edit} & \textbf{Evidence of completion} & \textbf{Recovery model} & \textbf{Scope} \\
\midrule
\multicolumn{6}{@{}l}{\textit{Plan- and execution-centric oversight}} \\
Plover \citep{venkatesan2026plover} & A structured plan over executable steps, locally revisable & Replanning of upcoming steps; executed history stays immutable & Executor step outcomes plus terminal screenshot check & Retry of pending steps within plan scope & Desktop GUI environments \\
\addlinespace
Sidekick \citep{chang2026sidekick} & Concurrent agent tracks and their status & Steering of which track continues and how & Agent-reported status indicators & Resumption of suspended tracks & Multi-task workflow context \\
\addlinespace
AgentLens \citep{kim2026agentlens} & An adaptive visual stream of the mobile UI & Which view renders and which touches are forwarded & Native UI layout inspection & Not part of the reported model & Mobile OS application layer \\
\addlinespace
ALLOY \citep{li2025alloy} & A reusable workflow graph & Which workflow is replayed and with what bindings & Per-step execution assertions & Workflow modification; state rollback not in the reported model & Web browser DOM tasks \\
\midrule
\multicolumn{6}{@{}l}{\textit{Malleable and generative views}} \\
DuetUI \citep{xu2026duetui} & A shared task context rendered as generative cards & Content of generated forms and mock services & Structural schema validation of generated UI & Session-level dialog reset & Generative UI task surfaces \\
\addlinespace
Jelly \citep{cao2025jelly} & An evolving data model rendered as a web application & The data model and UI schema behind the page & Schema conformance and semantic evaluation & In-memory rollback of the generated UI & Generative standalone apps \\
\addlinespace
Orca \citep{jiang2026orca} & Web components extracted into a mashup page & Arrangement of components across pages & Element-level DOM mutation observers & Global undo of local DOM modifications & Cross-page web DOM mashups \\
\midrule
\multicolumn{6}{@{}l}{\textit{Declarative target states}} \\
DMI \citep{wang2026dmi} & Declarative state targets over desktop widgets & Target states the interface should satisfy & Constraint evaluation and retry heuristics via the accessibility tree & Local retry; cross-app compensation not in the reported model & Windows desktop accessibility APIs \\
\midrule
\multicolumn{6}{@{}l}{\textit{State-contract governance (this work)}} \\
\textbf{TaskVM} & \textbf{A task contract over application facts: desired, preserve, observed} & \textbf{A checkable obligation on the world: what must change and what must remain untouched} & \textbf{Independent visual verification of fresh application observations, including non-interference} & \textbf{Typed reversibility over real GUI effects: reversible, compensatable, or irreversible with explicit locks} & \textbf{Multi-app web, mobile, and desktop} \\
\bottomrule
\end{tabular}
\end{table*}

\section{Stage 2 Interface Conditions}
\label{app:stage2-interfaces}

\begin{figure*}[t]
\centering
\includegraphics[width=0.85\textwidth, trim=26 225 25 218, clip]{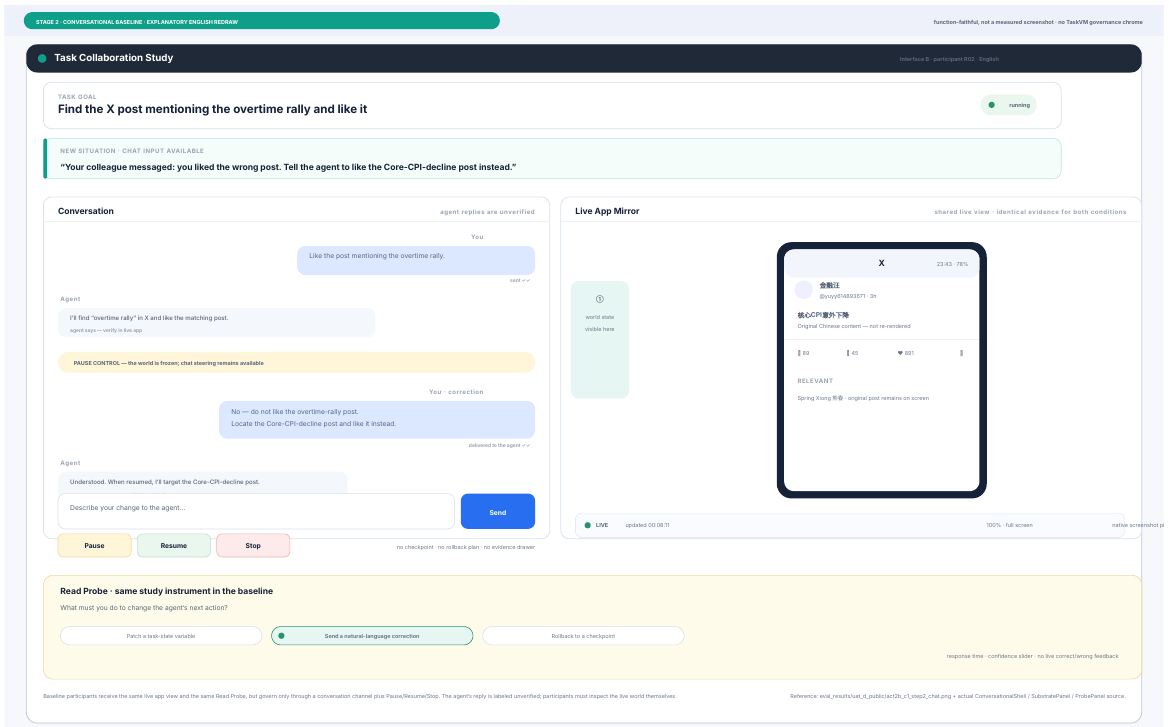}
\caption{Stage 2 baseline Conversational CUA interface: an application screencast overlaid with a natural language chat drawer. Due to language localization requirements, the interface has been semantically redrawn in the target language.}
\label{fig:interface-conv}
\vspace{1.5em}
\includegraphics[width=0.85\textwidth, trim=26 225 25 218, clip]{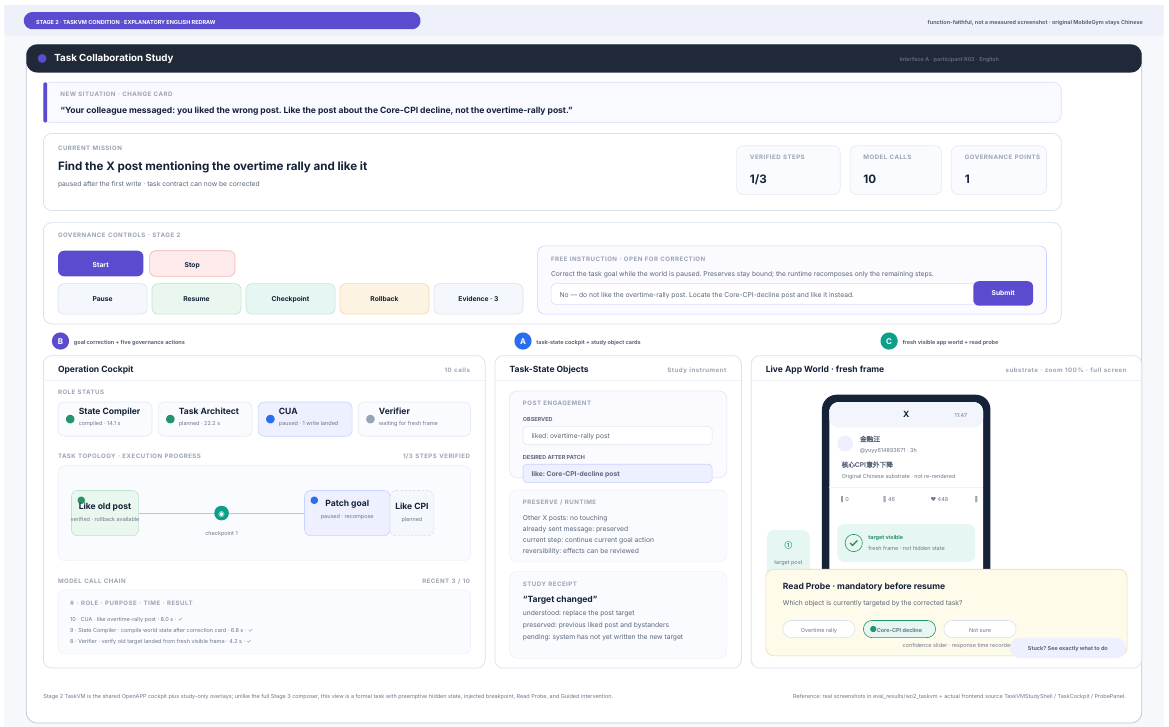}
\caption{Stage 2 TaskVM interface: the governance cockpit (Figure~\ref{fig:interface}) with the live application view and progress panels beside the editable task-state controls. Due to language localization requirements, the interface has been semantically redrawn in the target language.}
\label{fig:interface-taskvm}
\end{figure*}

The two interface conditions compared in Stage 2 are shown separately below.
In the baseline Conversational CUA condition (Figure~\ref{fig:interface-conv}), participants supervise an agent through an application screencast accompanied by an interactive chat drawer.
In the TaskVM condition (Figure~\ref{fig:interface-taskvm}), participants govern execution through the governance cockpit with the live application view beside the editable task-state controls.

\section{Per-Family COV Outcomes}
\label{app:stage1-families}

\begin{figure*}[t]
\centering
\includegraphics[width=0.9\textwidth]{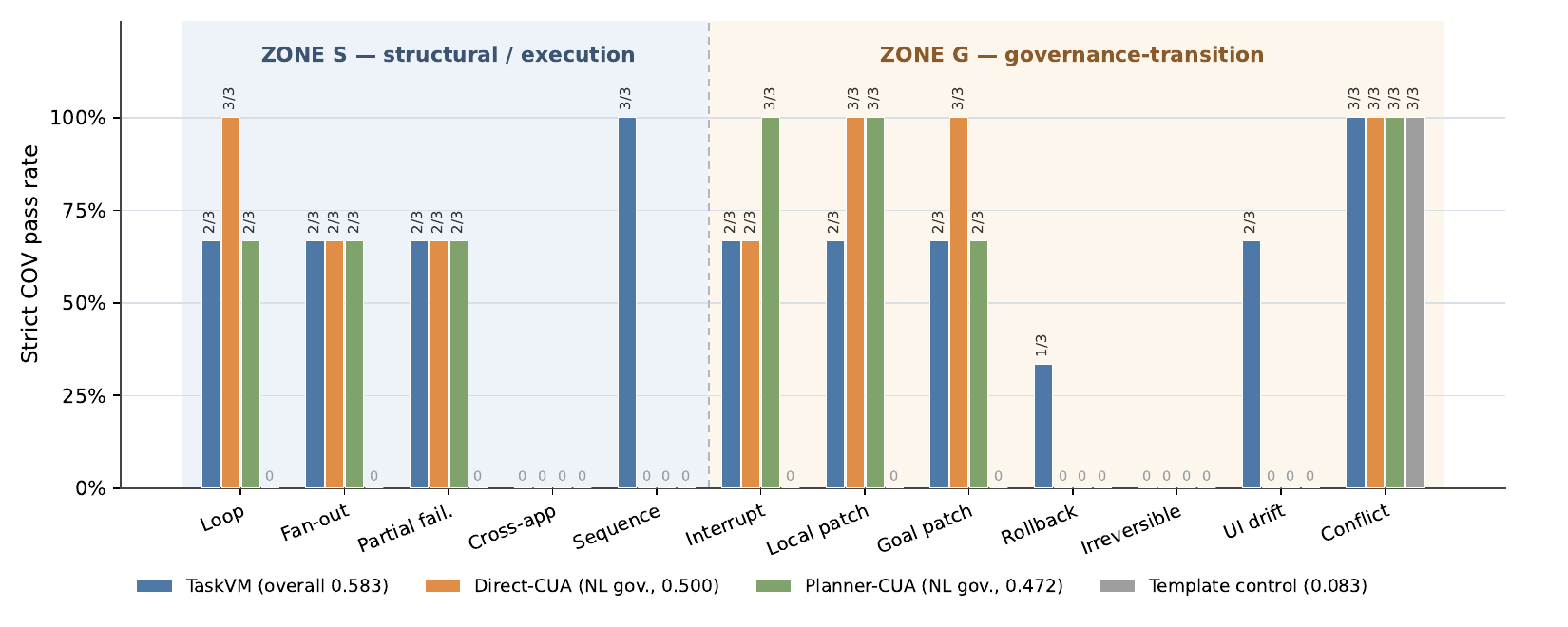}
\caption{Per-family strict COV pass rates across conditions on the frozen suite (36 instances in 12 structural families), grouped into Zone-S (structural and execution families) and Zone-G (governance-dense families).}
\label{fig:stage1-families}
\end{figure*}

Figure~\ref{fig:stage1-families} reports the strict COV pass rate for every task family in the frozen benchmark suite.
Aggregate condition-level rates are the primary comparison (Figure~\ref{fig:stage1-results}); per-family rates are noisy by construction, since each cell aggregates only a handful of trials, and are reported here for completeness and auditability.

\section{Stage 1 Per-Instance Outcomes}
\label{app:s1-instances}

Table~\ref{tab:s1-instances} lists the strict COV verdict of every frozen instance in the skills-OFF main wave, so that the aggregate rates in Table~\ref{tab:main-comparison} can be audited cell by cell.
A~\ding{51}{} means every applicable predicate passed on that instance's trial (Section~\ref{app:predicates}).
The Template Control column is omitted: it is an invalid sanity-check harness rather than a baseline (Section~\ref{app:stage1-honesty}); its per-instance record is preserved in the frozen data.
All four families in which no condition passed (cav, irv, and rbv variants rbv-02/03) are retained; no failing family is hidden.

\begin{table*}[!t]
\centering
\caption{Per-instance strict COV outcomes in the frozen skills-OFF wave (36 instances in 12 families), using the same family-to-zone dictionary as Figure~\ref{fig:stage1-families}. Zone-S (structural/execution): Loop \texttt{blv}, Fan-out \texttt{fav}, Partial-failure \texttt{pfv}, Cross-app \texttt{cav}, Sequence \texttt{sqv}. Zone-G (governance-dense): Conflict \texttt{cfv}, UI-drift \texttt{drv}, Interrupt \texttt{itv}, Local-patch \texttt{lpv}, Goal-patch \texttt{gpv}, Irreversible \texttt{irv}, Rollback \texttt{rbv}. Totals: TaskVM 21/36, Direct-CUA 18/36, Planner-CUA 17/36.}
\label{tab:s1-instances}
\scriptsize
\setlength{\tabcolsep}{3pt}
\begin{tabular}{@{}llccc@{\hspace{0.8em}}llccc@{}}
\toprule
\textbf{Family / Instance} & \textbf{Zn} & \textbf{TM} & \textbf{DC} & \textbf{PC} & \textbf{Family / Instance} & \textbf{Zn} & \textbf{TM} & \textbf{DC} & \textbf{PC} \\
\midrule
\multicolumn{5}{@{}l}{\textit{Zone-S (structural / execution)}} & \multicolumn{5}{l}{\textit{Zone-G (governance-dense), continued}} \\
\textit{Loop} \texttt{blv-01}         & S & \ding{51} & \ding{51} & \ding{51} & \textit{UI drift} \texttt{drv-02}      & G & -- & -- & -- \\
\textit{Loop} \texttt{blv-02}         & S & -- & \ding{51} & \ding{51} & \textit{UI drift} \texttt{drv-03}      & G & \ding{51} & -- & -- \\
\textit{Loop} \texttt{blv-03}         & S & \ding{51} & \ding{51} & -- & \textit{Interrupt} \texttt{itv-01}          & G & \ding{51} & \ding{51} & \ding{51} \\
\textit{Fan-out} \texttt{fav-01}      & S & \ding{51} & \ding{51} & \ding{51} & \textit{Interrupt} \texttt{itv-02}          & G & \ding{51} & \ding{51} & \ding{51} \\
\textit{Fan-out} \texttt{fav-02}      & S & -- & -- & -- & \textit{Interrupt} \texttt{itv-03}          & G & -- & -- & \ding{51} \\
\textit{Fan-out} \texttt{fav-03}      & S & \ding{51} & \ding{51} & \ding{51} & \textit{Local patch} \texttt{lpv-01}        & G & \ding{51} & \ding{51} & \ding{51} \\
\textit{Partial fail.} \texttt{pfv-01} & S & \ding{51} & \ding{51} & \ding{51} & \textit{Local patch} \texttt{lpv-02}       & G & \ding{51} & \ding{51} & \ding{51} \\
\textit{Partial fail.} \texttt{pfv-02} & S & -- & -- & -- & \textit{Local patch} \texttt{lpv-03}        & G & -- & \ding{51} & \ding{51} \\
\textit{Partial fail.} \texttt{pfv-03} & S & \ding{51} & \ding{51} & \ding{51} & \textit{Goal patch} \texttt{gpv-01}        & G & \ding{51} & \ding{51} & \ding{51} \\
\textit{Cross-app} \texttt{cav-01}    & S & -- & -- & -- & \textit{Goal patch} \texttt{gpv-02}           & G & -- & \ding{51} & \ding{51} \\
\textit{Cross-app} \texttt{cav-02}    & S & -- & -- & -- & \textit{Goal patch} \texttt{gpv-03}           & G & \ding{51} & \ding{51} & -- \\
\textit{Cross-app} \texttt{cav-03}    & S & -- & -- & -- & \textit{Irreversible} \texttt{irv-01}         & G & -- & -- & -- \\
\textit{Sequence} \texttt{sqv-01}     & S & \ding{51} & -- & -- & \textit{Irreversible} \texttt{irv-02}         & G & -- & -- & -- \\
\textit{Sequence} \texttt{sqv-02}     & S & \ding{51} & -- & -- & \textit{Irreversible} \texttt{irv-03}         & G & -- & -- & -- \\
\textit{Sequence} \texttt{sqv-03}     & S & \ding{51} & -- & -- & \textit{Rollback} \texttt{rbv-01}             & G & \ding{51} & -- & -- \\
\midrule
\multicolumn{5}{@{}l}{\textit{Zone-G (governance-dense)}} & \textit{Rollback} \texttt{rbv-02}             & G & -- & -- & -- \\
\textit{Conflict} \texttt{cfv-01}     & G & \ding{51} & \ding{51} & \ding{51} & \textit{Rollback} \texttt{rbv-03}             & G & -- & -- & -- \\
\textit{Conflict} \texttt{cfv-02}     & G & \ding{51} & \ding{51} & \ding{51} &  &  &  &  &  \\
\textit{Conflict} \texttt{cfv-03}     & G & \ding{51} & \ding{51} & \ding{51} &  &  &  &  &  \\
\textit{UI drift} \texttt{drv-01}     & G & \ding{51} & -- & -- &  &  &  &  &  \\
\bottomrule
\end{tabular}
\end{table*}

\section{Stage 2 Task Battery and Mid-Task Events}
\label{app:s2-battery}

Table~\ref{tab:s2-battery} specifies the full Stage 2 battery: four counterbalanced task pairs, each pair sharing one governance archetype across two surfaces.
The mid-task event was delivered by the experimenter at the trigger point shown, under the wizard-of-Oz replay protocol; the same event sequence, timing, and error manifestations were replayed in both interface conditions.
Each participant completed three of the eight tasks per condition; read/write probes were assigned to task slots by a deterministic balanced schedule (per-arm probe types per participant are recorded in the session logs).
All contact names are fictional entities of the offline simulator.

\begin{table*}[!t]
\centering
\caption{Stage 2 task battery (eight tasks, four pairs). Trigger points were determined by normalized execution events, not fixed wall-clock delays. Goal texts translated from the Chinese study script.}
\label{tab:s2-battery}
\scriptsize
\setlength{\tabcolsep}{3pt}
\begin{tabular}{@{}ll p{4.4cm} p{5.4cm} c@{}}
\toprule
\textbf{Task} & \textbf{Archetype} & \textbf{Participant goal [translated]} & \textbf{Mid-task event (trigger)} & \textbf{Budget} \\
\midrule
\texttt{r8-t1a} & pause--retarget & Social feed: like the post mentioning the ``overtime mobilization meeting''. & Experimenter announces the target changed to the ``core CPI surprise drop'' post; a similar decoy post exists (bound, pre-first-effect). & 4 min \\
\texttt{r8-t1b} & pause--retarget & Task board: mark ``organize the smoke-test checklist'' as done. & Target changes to ``update dependency list'' (same trigger). & 4 min \\
\texttt{r8-t2a} & mixed rollback & Social feed: like \emph{and} bookmark the overtime-meeting post. & After the first effect lands, the experimenter requests an undo; the interface must display the reversibility boundary. & 5 min \\
\texttt{r8-t2b} & mixed rollback & Chat: send a contact the message ``weekly meeting moved to Thursday 2\,pm''. & After send confirmation, undo is requested; a sent message is irreversible and must be honestly reported as such. & 5 min \\
\texttt{r8-t3a} & external drift & Social feed: like the ``team-building photo collection'' post. & World-side drift is injected after target binding, before the effect lands; the participant must reconcile the conflict. & 4 min \\
\texttt{r8-t3b} & external drift & Calendar: change the ``daily standup'' RSVP from \emph{accept} to \emph{tentative}. & Same drift injection (bound, pre-effect). & 4 min \\
\texttt{r8-t4a} & content drift, cross-app & Feed + chat: bookmark the overtime post \emph{and} relay a message to a contact that everyone must attend. & The post's content drifts before the relayed message lands, making the relayed claim stale; correction required. & 6 min \\
\texttt{r8-t4b} & content drift, cross-app & Feed + chat: bookmark the CPI post \emph{and} relay a risk-tracking message to a contact. & Same content-drift mechanic. & 6 min \\
\bottomrule
\end{tabular}
\end{table*}

\section{Stage 3 Usage in Aggregate}
\label{app:s3-catalogue}

To protect participant privacy, the anonymized archive retains per-episode runtime counters (goal formulation, status, model-call count) rather than complete GUI interaction traces, and some open-ended search terms typed by participants are withheld entirely.
We therefore report Stage 3 usage in aggregate (Table~\ref{tab:s3-usage}) instead of as per-task transcripts, and point readers to the open-ended responses (Table~\ref{tab:s3-qual}) for what participants themselves chose to disclose about their tasks.
Per-episode success is not adjudicated (Stage 3 embedded no probes and no per-task ground truth).
Model calls are the LLM invocations attributed to an episode by the runtime ledger; application and contact names appearing anywhere in the study are fictional entities of the offline simulator.

Diversity is visible at the aggregate level. The fourteen participants differ markedly in how they used the open session: total model calls span 3 to 149 per participant (median 52.5), episodes span 1 to 4 (median 3), and tasks touched five distinct application surfaces across the cohort (messaging: 11 participants; social feed: 7; video platform: 4; music: 3; content community: 2).
An independence audit confirmed that no two participants produced an identical episode sequence: even where two participants issued a textually identical goal, the runs differ in status, timing, and model-call counts (e.g., 3, 41, 46, 50, 63, and 73 calls on six runs of one shared formulation).

\begin{table}[t]
\centering
\caption{Stage 3 usage per returning participant, as retained in the anonymized archive. \#Form.\ = distinct goal formulations issued.}
\label{tab:s3-usage}
\footnotesize
\begin{tabular}{@{}lrrr r@{}}
\toprule
\textbf{PID} & \textbf{Episodes} & \textbf{Model calls} & \textbf{\#Apps} & \textbf{\#Form.} \\
\midrule
P01 & 4 & 74 & 2 & 4 \\
P02 & 3 & 52 & 2 & 2 \\
P05 & 3 & 9 & 2 & 2 \\
P06 & 4 & 42 & 1 & 3 \\
P10 & 3 & 93 & 3 & 3 \\
P13 & 3 & 149 & 3 & 2 \\
P20 & 3 & 93 & 3 & 2 \\
P21 & 3 & 41 & 1 & 3 \\
P22 & 1 & 3 & 1 & 1 \\
P25 & 4 & 143 & 2 & 3 \\
P27 & 1 & 3 & 1 & 1 \\
P30 & 3 & 111 & 1 & 3 \\
P35 & 2 & 39 & 3 & 2 \\
P37 & 2 & 53 & 2 & 2 \\
\midrule
\textbf{Cohort} & \textbf{39} & \textbf{905} & \textbf{5 surfaces} & \textbf{median 2} \\
\bottomrule
\end{tabular}
\end{table}

\section{Stage 3 Open-Ended Response Evidence Map}
\label{app:s3-qual}

Table~\ref{tab:s3-qual} maps every theme reported in Section~\ref{sec:stage3} to its supporting participants and a representative verbatim quote.
Themes were coded by the research team from the two open-ended questionnaire items; frequencies count distinct participants, a participant may contribute to several themes, and quotes are translated from the Chinese originals with the participant ID preserved.
We report open-ended responses, not interviews; no inter-rater reliability is claimed.

\begin{table*}[!t]
\centering
\caption{Stage 3 open-ended response evidence map. F = favourable theme, L = limitation theme. [translated] from Chinese originals.}
\label{tab:s3-qual}
\footnotesize
\begin{tabular}{@{}lp{1.9cm}lp{7.7cm}@{}}
\toprule
\textbf{Theme} & \textbf{Content} & \textbf{Participants} & \textbf{Representative quote [translated]} \\
\midrule
F1 & Faithful execution and precise intent understanding & P21, P37, P27, P25 (4) & ``The system recognized my intent and made the keyword changes, and it did not over-guess what I wanted to do.'' (P27) \\
F2 & Governance features used and valued mid-flight & P35, P30 (2) & ``Mid-execution I added an instruction to `ask Situ Nian whether she can work overtime'; the agent finished the original task, then correctly found the contact and surfaced the prompt. I found this delightful.'' (P30) \\
F3 & Participants spontaneously request stronger governance automation & P30, P27 (2) & ``Automatically set a checkpoint at each node, so I can roll back at any time.'' (P30); ``I hope the system can autonomously identify the critical points where problems may occur.'' (P27) \\
\midrule
L1 & Offline simulator has finite data coverage & P35, P06, P13 (3) & ``I wanted to search for Valorant, cats, and the `Blue-and-White Porcelain' MV, but the world model had none of them, which was quite frustrating.'' (P13) \\
L2 & Executor (model) reliability, orthogonal to the interface & P10, P20, P13, P22 (4) & A complex task ``fell into a loop of repeated comments'' (P20); repeated retries of the same send action (P13). \\
L3 & UI flow and tutorial coverage & P06, P01, P05, P20, P13, P25, P02 (7) & Progress not visible during retries (P06); features discovered too late, i.e., tutorial coverage rather than missing capability (P01, P05). \\
\bottomrule
\end{tabular}
\end{table*}

\section{Stage 3 Participant-Submitted Screenshots}
\label{app:s3-screenshots}

\begin{figure*}[!t]
\centering
\includegraphics[width=\textwidth]{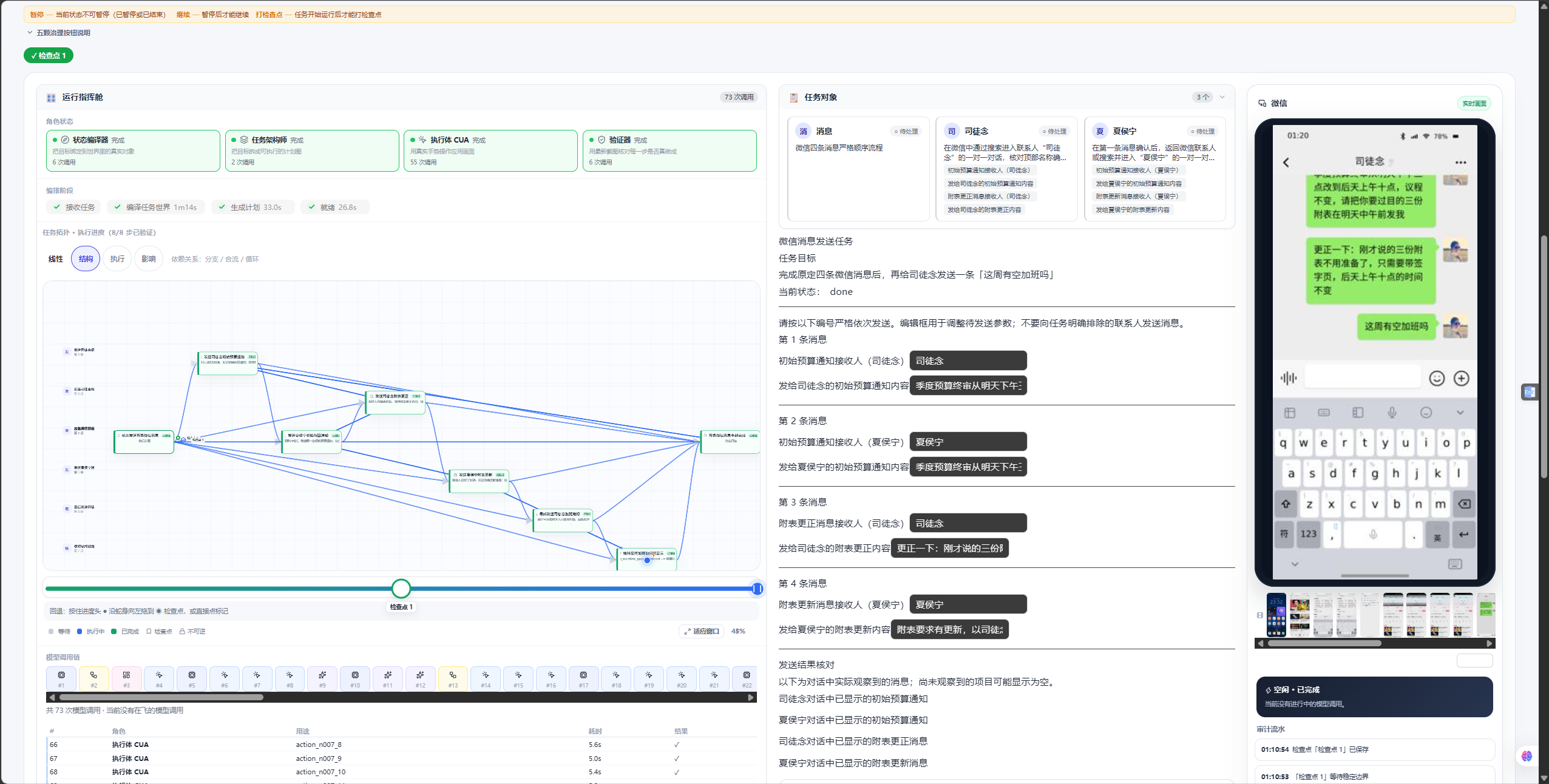}
\caption{Stage 3 cockpit screenshot voluntarily submitted by participant P30 immediately after a self-directed session.
The task was an ordered four-message chain in the simulated messaging application: rescheduling a quarterly budget review to two designated contacts, followed by a mid-chain correction. After the chain, the participant appended a self-authored final instruction asking one contact whether she could work overtime that week.
The left cockpit shows all four plan nodes passed and the goal status \emph{done} (73 model calls, matching the session ledger); the right panel shows the delivered messages inside the simulated app, including the appended overtime question, illustrating that a mid-flight user instruction was honored only after the original ordered chain had completed.
\textcolor{red}{Every personal name visible in this screenshot (e.g., the contact ``Situ Nian'') is a programmatically injected fictional identity of the offline simulator's seeded corpus; it does not denote any real person, any resemblance to a real individual is purely coincidental, and its appearance does not constitute a failure of anonymization.}}
\label{fig:s3-screenshot-msg}
\end{figure*}

\begin{figure*}[!t]
\centering
\includegraphics[width=0.62\textwidth]{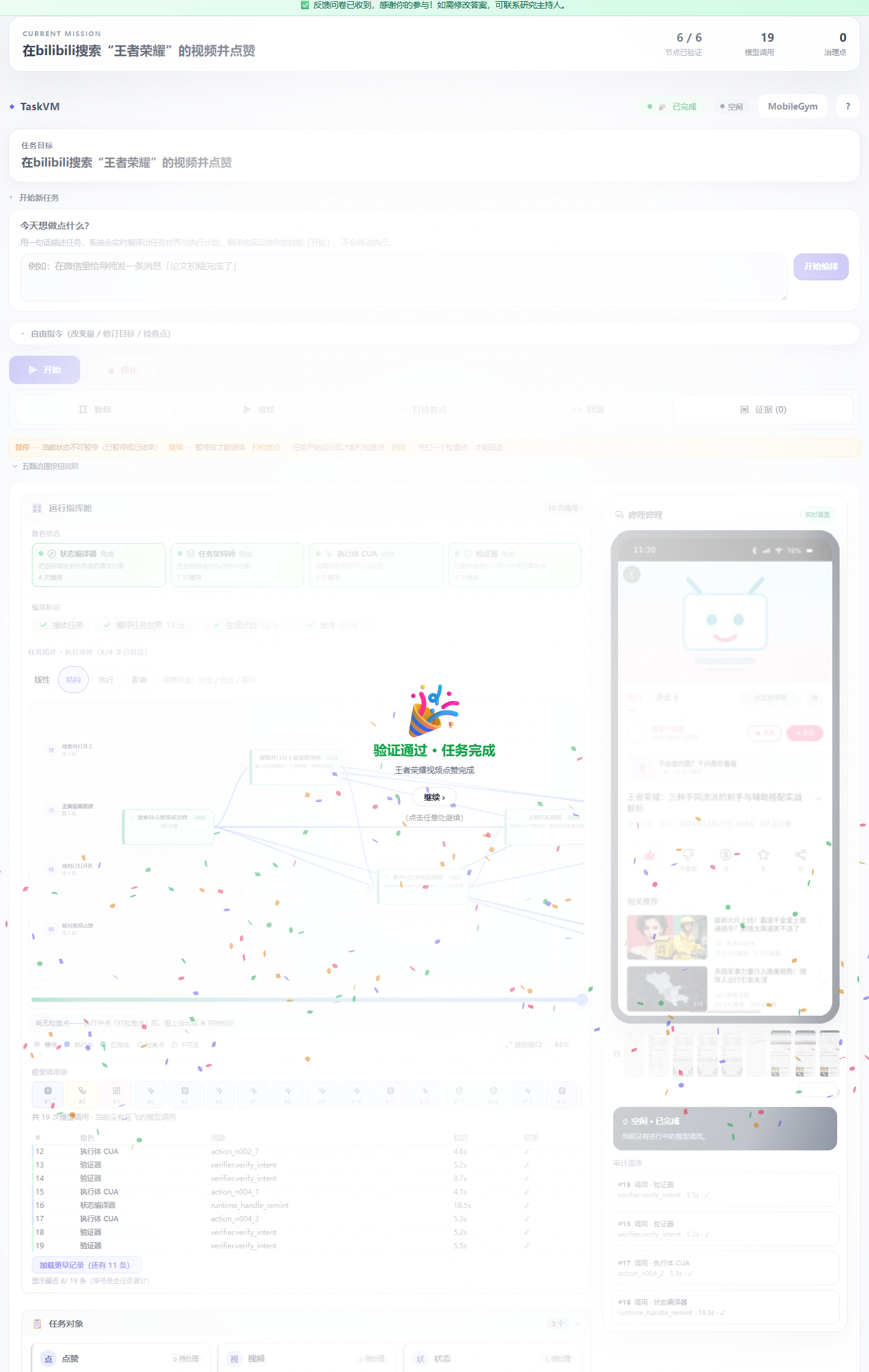}
\caption{Stage 3 screenshot voluntarily submitted by participant P06 at the moment the system reported task completion.
The participant authored the task ``search Bilibili for an \emph{Arena of Valor} video and like it''; the image captures the verification-pass state (banner ``Verification passed, task completed''; all six plan nodes verified, 19 model calls, zero governance interventions used), while the right panel shows the liked video inside the simulated video application.
\textcolor{red}{Any personal name that may appear in this screenshot is a programmatically injected fictional identity of the offline simulator's seeded corpus; it does not denote any real person, any resemblance to a real individual is purely coincidental, and its appearance does not constitute a failure of anonymization.}}
\label{fig:s3-screenshot-like}
\end{figure*}

Figures~\ref{fig:s3-screenshot-msg} and~\ref{fig:s3-screenshot-like} were spontaneously captured and voluntarily submitted to the research team by two Stage 3 participants.
The interface language of the study deployment is Chinese; both images are reproduced \textbf{without any modification or translation}.

\section{Stage 2 Measurement Instruments}
\label{app:instruments}

Table~\ref{tab:instruments} reproduces every item of the condition-end questionnaire administered once per condition (instrument \texttt{condition\_end\_v1}, Chinese administration with officially paired English strings; English shown verbatim).
Composites: NASA-TLX = unweighted mean of the six 0--100 sliders (the performance slider runs Perfect\,$\to$\,Failure, so all dimensions share the demand direction); UMUX-Lite = $\mathrm{mean}(\mathrm{items}-1)/6\times100$; recovery control = mean of the four \texttt{rc\_*} items; trust = mean of \texttt{tia\_trust} and \texttt{tia\_rely}; perceived transparency = mean of the three predictability items with \texttt{tia\_unpredictable\_R} reverse-scored ($8-\mathrm{raw}$); support satisfaction = mean of the three \texttt{asq\_*} items.
All Likert items used 1 (\emph{Strongly disagree}) to 7 (\emph{Strongly agree}).
The post-study background questionnaire (Appendix~\ref{app:demographics}) is a separate instrument.

Read probes were embedded in drift-triggered trials: three multiple-choice questions per probe covering committed effects, the effective goal, and governance facts, drawn from a frozen probe bank (per-task options and answer keys frozen before data collection; scoring = exact key match per question), preceded by a single 0--100 confidence rating for the whole probe.
Write probes scored the participant's own intervention at the scripting boundary and at trial end against the frozen expectation for that task (both endpoints recorded per participant per arm).

\begin{table*}[!t]
\centering
\caption{Condition-end questionnaire: all administered items (English strings verbatim).}
\label{tab:instruments}
\scriptsize
\setlength{\tabcolsep}{3pt}
\begin{tabular}{@{}llp{8.9cm}l@{}}
\toprule
\textbf{Module} & \textbf{Item} & \textbf{Text} & \textbf{Scale} \\
\midrule
NASA-TLX & \texttt{tlx\_mental} & How mentally demanding was the task? & 0--100 \\
 & \texttt{tlx\_physical} & How physically demanding was the task? & 0--100 \\
 & \texttt{tlx\_temporal} & How hurried or rushed was the pace of the task? & 0--100 \\
 & \texttt{tlx\_performance} & How successful were you in accomplishing what you were asked to do? & 0--100 (Perfect$\to$Failure) \\
 & \texttt{tlx\_effort} & How hard did you have to work to accomplish your level of performance? & 0--100 \\
 & \texttt{tlx\_frustration} & How insecure, discouraged, irritated, stressed, and annoyed were you? & 0--100 \\
\midrule
UMUX-Lite & \texttt{umux\_capability} & This system's capabilities meet my requirements. & 1--7 \\
 & \texttt{umux\_ease} & This system is easy to use. & 1--7 \\
\midrule
Recovery control (study-specific) & \texttt{rc\_influence} & I felt able to influence what the system was doing during execution. & 1--7 \\
 & \texttt{rc\_stop\_redirect} & I knew how to stop or redirect the system when needed. & 1--7 \\
 & \texttt{rc\_correct} & When the system did something I did not want, I was able to correct it. & 1--7 \\
 & \texttt{rc\_overall} & Overall, I felt in control of how the task was carried out. & 1--7 \\
\midrule
TIA predictability subset & \texttt{tia\_state\_clear} & The system state was always clear to me. & 1--7 \\
 & \texttt{tia\_unpredictable\_R} & The system reacts unpredictably. & 1--7 (reversed) \\
 & \texttt{tia\_understand\_why} & I was able to understand why things happened. & 1--7 \\
\midrule
ASQ (adapted, per condition block) & \texttt{asq\_ease} & I am satisfied with the ease of completing the tasks in this block. & 1--7 \\
 & \texttt{asq\_time} & I am satisfied with the time it took to complete the tasks in this block. & 1--7 \\
 & \texttt{asq\_support} & I am satisfied with the information and support the system provided while completing this block. & 1--7 \\
\midrule
TIA trust & \texttt{tia\_trust} & I trust the system. & 1--7 \\
 & \texttt{tia\_rely} & I can rely on the system. & 1--7 \\
\bottomrule
\end{tabular}
\end{table*}

\section{Evaluation Configuration}
\label{app:evalconfig}

\begin{table*}[!t]
\centering
\caption{Model and runtime configuration by stage and condition. All LLM roles used the same production model; differences are architectural, not model-level.}
\label{tab:evalconfig}
\scriptsize
\setlength{\tabcolsep}{3pt}
\begin{tabular}{@{}lp{5.1cm}p{6.2cm}@{}}
\toprule
\textbf{Condition} & \textbf{Model roles} & \textbf{Notes} \\
\midrule
\multicolumn{3}{@{}l}{\textit{Stage 1 (automated benchmark, frozen waves)}} \\
TaskVM & GPT-5.6-sol $\times$ 4 roles: state compiler, task architect, CUA executor, independent model verifier & Verifier reads fresh screenshots only; per-trial role call ledgers retained \\
Direct-CUA & GPT-5.6-sol, single agentic loop & Same seeds, same substrate \\
Planner-CUA & GPT-5.6-sol, planner then executor & Same seeds, same substrate \\
Template Control & none (scripted gesture driver) & Assembly sanity check; zero LLM calls \\
\midrule
\multicolumn{3}{@{}l}{\textit{Stage 2 (user study)}} \\
Both arms & replay rig (no live model) & Identical recorded trajectory + world events + error manifestation in both arms \\
\midrule
\multicolumn{3}{@{}l}{\textit{Stage 3 (open exploration)}} \\
TaskVM & GPT-5.6-sol, same role pipeline as Stage 1 & Live model; participants' self-issued tasks \\
\bottomrule
\end{tabular}
\end{table*}

In all stages the user-facing runtime observes only rendered pixels and emits standard GUI gestures; oracle-only capabilities (state resets, internal-state reads, deterministic event injection) are confined to an isolated evaluation harness and are never exposed to the interactive session, as detailed in the prototype section.
Per-trial manifests record model identity, per-role call counts, and wall time; frozen artifacts per instance are retained in the consolidated archive cited in Section~\ref{sec:stage1}.


\section{Participant Demographics and Background}
\label{app:demographics}

\subsection{What was collected, and from whom}

Demographic and background information was collected after the main study through two voluntary channels.
(i)~At compensation time, participants could provide identifying contact information together with self-reported gender, age, and affiliation; 36 of the 37 completers did so, and one (P08) declined all background collection.
(ii)~On 2026-09-12 we additionally administered a five-item background questionnaire (field of study; professional background related to AI agents; usage frequency of three classes of AI systems; AI products used; purposes of use) as a web page gated by roster name verification; 35 of the 37 completers answered it, one (P32) submitted compensation information but declined the questionnaire.
The questionnaire was administered in Chinese; a faithful English transcription is included in the supplementary material, and Table~\ref{tab:instrument-map} maps the dataset's option codes to the original wording.
No information in this appendix was inferred from names, photos, or researcher impressions; undeclared fields are reported as such.

\subsection{Aggregate profile}

Table~\ref{tab:demo-summary} summarizes the sample.
Participants' self-reported affiliations covered universities, public research institutes, and industry (32 academic, 4 industry among the 36 on record); participants reported their own gender (15 female, 21 male, 1 undisclosed) and age ($M{=}24.7$, $\mathit{SD}{=}4.3$, range 18--42).

\paragraph{Why field labels alone do not make this an ``insider'' sample.}
A majority of participants (20/35) classified themselves into CS/AI disciplines, but the questionnaire deliberately separated disciplinary labels from agent-specific expertise, and the two diverge sharply.
Most CS/AI-classified participants specialize in areas whose day-to-day work does not involve building or evaluating GUI agents, such as computer vision, embodied AI, computer architecture, NLP post-training, signal processing, and similar fields.
Asked directly about \emph{professional} background with agents, only 2/35 (5.7\%) reported direct experience with computer-use/GUI agents, 5/35 reported agent (but not GUI-agent) backgrounds, 12/35 reported AI work unrelated to agents, and 16/35 reported no or only limited professional exposure to AI agents at all.
Usage told the same story: 14/35 (40\%) had never used a GUI agent or had tried one only once or twice before the study.
The sample is therefore best characterized as technically literate but predominantly \emph{novice with respect to GUI agents}, the population for which governance interfaces must work.

Participants' existing AI usage was mainstream rather than specialist.
The most-used products were conversational assistants (ChatGPT 31/35, DeepSeek 24, Claude 19, Gemini 18); coding assistants were common (Claude Code 13, GitHub Copilot 6, other coding agents 6, Cursor 3), while research-style tools (6), general-purpose agents (5), and any standalone computer-use/GUI-agent product (only 2) were rare.
Reported purposes centered on research (27/35), coding (16), learning (15), and writing (10); delegating a multi-step task to an agent for autonomous completion was reported by only one participant.
One additional participant reported essentially no AI use.

\begin{table*}[!t]
\centering
\caption{Aggregate demographics and background of the Stage 2 sample ($N{=}37$). Gender, age, and affiliation from the compensation record ($n{=}36$); field and agent-related items from the background questionnaire ($n{=}35$).}
\label{tab:demo-summary}
\scriptsize
\setlength{\tabcolsep}{4pt}
\begin{tabular}{@{}llr@{}}
\toprule
\textbf{Dimension} & \textbf{Category} & \textbf{$n$} \\
\midrule
Gender (self-reported) & Female / Male / Undisclosed & 15 / 21 / 1 \\
Age band & 18--24 / 25--34 / 35--44 & 19 / 15 / 2 \\
Affiliation sector & Academic / Industry & 32 / 4 \\
\midrule
Field of study ($n{=}35$) & CS/AI disciplines & 20 \\
 & EE / basic sciences / other STEM & 9 \\
 & Business / humanities / other & 6 \\
\midrule
Agent professional background$^{\star}$ ($n{=}35$) & 1: none & 5 \\
 & 2: limited exposure & 11 \\
 & 3: AI-related, not agents & 12 \\
 & 4: AI agents, not GUI agents & 5 \\
 & 5: direct computer-use / GUI agents & 2 \\
\midrule
GUI-agent usage, past 6 months ($n{=}35$) & Never / Tried once or twice & 7 / 7 \\
 & $<$Monthly / Monthly & 4 / 4 \\
 & Weekly / Daily / Multiple times daily & 4 / 3 / 6 \\
\bottomrule
\multicolumn{3}{@{}l}{\scriptsize $^{\star}$Level wording per the questionnaire; see Table~\ref{tab:instrument-map}.}
\end{tabular}
\end{table*}

\subsection{Instrument mapping}

\begin{table*}[!t]
\centering
\caption{Mapping from dataset option codes to the original questionnaire wording (abridged English transcription; the full five-item instrument, including its warning boxes distinguishing GUI agents from other AI tools, is in the supplementary material).}
\label{tab:instrument-map}
\footnotesize
\begin{tabular}{@{}ll@{}}
\toprule
\textbf{Item / code} & \textbf{Original option wording (abridged)} \\
\midrule
Q1 \texttt{ai\_ml} & Artificial Intelligence / Machine Learning / Deep Learning \\
Q1 \texttt{cv} & Computer Vision / Image Processing \\
Q1 \texttt{nlp} & Natural Language Processing / Large Language Models \\
Q1 \texttt{robotics} & Robotics / Embodied Intelligence / Automation / Control \\
Q1 \texttt{hci} & Human--Computer Interaction / UX / Interaction Design \\
Q1 \texttt{systems} & Computer Systems / Operating Systems / Distributed Systems \\
Q1 \texttt{ee} & Electronics / Chips / Embedded Systems \\
Q1 \texttt{math} & Mathematics / Statistics / Physics and other basic sciences \\
Q1 \texttt{other\_eng} & Other engineering or STEM disciplines \\
Q1 \texttt{business} & Business / Management / Economics / Finance \\
Q1 \texttt{humanities} & Humanities / Social Sciences / Arts and Design \\
\midrule
Q2 level 1 & No relevant professional experience with AI/agent technologies \\
Q2 level 2 & Limited exposure (AI only as auxiliary or peripheral) \\
Q2 level 3 & AI-related professional background, not in agents \\
Q2 level 4 & Direct AI-agent background, not computer-use/GUI agents \\
Q2 level 5 & Direct computer-use / GUI-agent professional background \\
\midrule
Q3 rows & A: conversational LLMs; B: general AI agents (non-GUI); C: computer-use/GUI agents \\
Q3 scale & never, tried once or twice, $<$monthly, monthly, weekly, daily, multiple times daily \\
\bottomrule
\end{tabular}
\end{table*}

\subsection{Per-participant roster}

Table~\ref{tab:demo-roster} lists every completer's self-reported information.
Field categories follow the questionnaire's own option wording. Note in particular that the code \texttt{math} corresponded to ``Mathematics / Statistics / Physics and other basic sciences,'' so basic-science majors (e.g., physics) appear under that label; the parenthesized detail is the participant's own free-text specialization, translated.
Category codes follow the raw submission log; for one participant (P13) the compensation copy and the submission log differ on the category code (\texttt{other} vs.\ \texttt{math}), and we report the submission log.

\begin{table*}[!t]
\centering
\caption{Per-participant demographics and background roster ($N{=}37$). Age = participant's band from the exact self-reported age; Sector: Acad.\ = university or public research institute, Ind.\ = industry employer (institution names withheld against re-identification); Field = questionnaire category with the participant's free-text specialization in parentheses [translated]; AgentBG = agent professional background level (Table~\ref{tab:instrument-map}); GUI-use = self-reported computer-use/GUI-agent usage frequency over the past 6 months; S3 = returned for Stage 3.}
\label{tab:demo-roster}
\scriptsize
\begin{tabular}{@{}llllp{4.9cm}clc@{}}
\toprule
\textbf{PID} & \textbf{Gender} & \textbf{Age} & \textbf{Sector} & \textbf{Field of study (specialization)} & \textbf{AgentBG} & \textbf{GUI-use} & \textbf{S3} \\
\midrule
P01 & F & 25--34 & Ind. & AI / ML / DL (computer vision) & 2 & Daily & \ding{51} \\
P02 & M & 25--34 & Acad. & AI / ML / DL (computer vision) & 3 & Multi-daily & \ding{51} \\
P03 & M & 18--24 & Acad. & NLP / LLMs (opponent modeling) & 4 & Multi-daily &  \\
P04 & F & 18--24 & Acad. & HCI / UX (AI--human interaction) & 4 & Tried &  \\
P05 & M & 25--34 & Acad. & Computer Systems (computer architecture) & 3 & Tried & \ding{51} \\
P06 & M & 18--24 & Acad. & Math.\ / Physics / Basic Sci.\ (systems analysis \& integration; data fusion) & 2 & Never & \ding{51} \\
P07 & F & 25--34 & Acad. & Other Eng.\ / STEM (SAR imaging algorithms) & 2 & Daily &  \\
P08 & \multicolumn{6}{l}{not disclosed (declined all background collection)} &  \\
P09 & F & 25--34 & Acad. & Humanities / Arts (journalism \& communication) & 2 & Monthly &  \\
P10 & F & 18--24 & Acad. & Humanities / Arts (advertising design) & 2 & Never & \ding{51} \\
P11 & M & 18--24 & Acad. & AI / ML / DL (sign-language translation) & 3 & Tried &  \\
P12 & M & 18--24 & Acad. & NLP / LLMs (AI safety) & 4 & Daily &  \\
P13 & F & 18--24 & Acad. & Math.\ / Physics / Basic Sci.\ (AI4Science, protein domain) & 3 & Tried & \ding{51} \\
P14 & F & 25--34 & Acad. & NLP / LLMs (LLM post-training) & 3 & Weekly &  \\
P15 & F & 25--34 & Acad. & AI / ML / DL (CS\&T, information processing) & 4 & Monthly &  \\
P16 & F & 18--24 & Acad. & Humanities / Arts (journalism, new media) & 2 & Tried &  \\
P17 & M & 18--24 & Acad. & Other Eng.\ / STEM (geographic information science) & 4 & Weekly &  \\
P18 & M & 18--24 & Acad. & NLP / LLMs (computational linguistics) & 3 & $<$Mo. &  \\
P19 & F & 18--24 & Acad. & AI / ML / DL (artificial intelligence) & 2 & $<$Mo. &  \\
P20 & M & 25--34 & Acad. & AI / ML / DL (embodied AI) & 3 & Tried & \ding{51} \\
P21 & M & 18--24 & Acad. & Computer Vision & 5 & Multi-daily & \ding{51} \\
P22 & M & 18--24 & Acad. & NLP / LLMs (multimodal LLMs) & 3 & Never & \ding{51} \\
P23 & F & 25--34 & Acad. & Business / Econ.\ (economics) & 2 & $<$Mo. &  \\
P24 & M & 35--44 & Acad. & Robotics / Embodied (embodied AI) & 1 & Multi-daily &  \\
P25 & M & 18--24 & Acad. & AI / ML / DL (pattern recognition \& intelligent systems) & 3 & Multi-daily & \ding{51} \\
P26 & M & 18--24 & Ind. & AI / ML / DL (GUI agents) & 5 & Multi-daily &  \\
P27 & M & 25--34 & Ind. & AI / ML / DL (pattern recognition \& intelligent systems) & 3 & Weekly & \ding{51} \\
P28 & F & 18--24 & Acad. & Humanities / Arts (applied psychology) & 2 & Never &  \\
P29 & M & 35--44 & Ind. & AI / ML / DL (embodied AI) & 3 & Monthly &  \\
P30 & M & 18--24 & Acad. & Computer Vision & 3 & $<$Mo. & \ding{51} \\
P31 & F & 25--34 & Acad. & Electronics / Embedded (electronic packaging) & 1 & Never &  \\
P32 & F & 25--34 & Acad. & \multicolumn{3}{l}{declined the background questionnaire} &  \\
P33 & M & 25--34 & Acad. & Electronics / Embedded (signal \& information processing) & 2 & Monthly &  \\
P34 & M & 18--24 & Acad. & Electronics / Embedded (electronic information eng.) & 1 & Tried &  \\
P35 & F & 18--24 & Acad. & Humanities / Arts (Law) & 1 & Never & \ding{51} \\
P36 & M & 25--34 & Acad. & Other Eng.\ / STEM (environmental engineering) & 1 & Never &  \\
P37 & M & 25--34 & Acad. & Math.\ / Physics / Basic Sci.\ (experimental particle physics) & 2 & Weekly & \ding{51} \\
\bottomrule
\end{tabular}
\end{table*}

\end{document}